\documentclass[journal,onecolumn]{IEEEtran}
\pdfoutput=1 
\usepackage{verbatim}

\usepackage{cite}

\usepackage{graphicx}
\usepackage{subfigure}
\usepackage{wrapfig}
\usepackage{enumerate}
\usepackage{color}
\usepackage{wrapfig}
\ifCLASSINFOpdf
\else
\fi
\usepackage{hyperref}
\usepackage{amsmath}

\usepackage{algorithm}
\usepackage{algpseudocode}
\usepackage{mathtools}

\usepackage{array}
\begin{document}

\pagenumbering{arabic}
%
\title{Variational Embedding Multiscale Sample Entropy: \\complexity-based analysis for multichannel systems}
%
%
%

 \author{Hongjian Xiao and Danilo P. Mandic}

\maketitle

\begin{abstract}
To quantify the complexity of a system, entropy-based methods have received considerable critical attention in real-world data analysis. Among numerous entropy algorithms, amplitude-based formulas, represented by Sample Entropy, suffer from a limitation of data length especially when it comes to practical scenarios. And this shortcoming is further highlighted by involving coarse graining procedure in multi-scale process. The unbalance between embedding dimension and data size will undoubtedly result in inaccurate and undefined estimation. To that cause, Variational Embedding Multiscale Sample Entropy is proposed in this paper, which assign signals from various channels with distinct embedding dimensions. And this algorithm is tested by both stimulated and real signals. Furthermore, the performance of the new entropy is investigated and compared with Multivariate Multiscale Sample Entropy and Variational Embedding Multiscale Diversity Entropy. Two real-world database, wind data sets with varying regimes and physiological database recorded from young and elderly people, were utilized. As a result, the proposed algorithm gives an improved separation for both situations.
\end{abstract}


%
\IEEEpeerreviewmaketitle

\section{Introduction}
%
%
%
%
\IEEEPARstart{T}{he} investigation of complexity of real world signals we have witnessed a boom in the last decades. Now recognized to have the same importance as the properties in the time and frequency domain, the complexity of a data set is a unique feature that can be utilized to understand a signal generating mechanism via nonlinear analytical tools. The studies of complexity have covered a wide spectrum from the fault diagnosis of rotating machine \cite{Ref48,Ref51,Ref56} through to the early detection of disease and sickness in humans \cite{Ref8,Ref37,Ref68,Ref81}. Indeed, bio signals exhibit high degrees of irregularities and complex dynamical behaviours \cite{Ref80}. Such nonlinear dynamics results from the interactions between human body (organisms) and peripheral environment and exhibits continuous fluctuations in time domain \cite{Ref79}. Complexity Loss Theory (CLT) states the potential relationship between the complexity of physical signals and health of an individual where the higher degree of complexity indicates healthier condition of the individual \cite{Ref78}. However, new developments have declared that pathology also exhibits an increase in complexity based on structure, that is, a decrease of self-correlated complexity will also be observed in healthy body \cite{Ref1}. 

Although the definition of structural complexity is inconsistent \cite{Ref27}, there are several commonly used methods to quantify the degree of dynamics where entropy-based methodologies are popular ones. Compared to other methods to estimate complexity of nonlinear systems, such as fractal dimension \cite{Ref80} and recurrence plots \cite{Ref79}, entropy analysis holds the advantages of simplicity and noise robustness \cite{Ref77}. The features of the loss of complexity (LOC) manifest themselves through, for example, an increase in randomness, less regularity, breakdown of long-term correlations, multiscale variability, and time irreversibility \cite{Ref27}. To this end, a large number of various entropy algorithms have been proposed to quantify the different features of complexity, or more precisely, the degree of complexity based on different definitions.

The two early entropy algorithms that have been widely used are the Approximate Entropy (ApEn) \cite{Ref84} and Sample Entropy (SampEn) \cite{Ref22}, proposed in 1991 and 2000 respectively. Modified by ApEn, SampEn is able to reduce the bias experienced by ApEn by removing self-matching and simultaneously has less dependency on the data length giving relatively higher consistency \cite{Ref22}. Both ApEn and SampEn were developed to quantify the randomness and irregularity of the system. Generally speaking, the lower the value of SampEn, the less complex the system. However, truly complex signals exhibit varying structures across multiple time scales, while long-range correlations fail to be observed by single-scale Sample Entropy analysis. To this end, Costa \textit{et al}. introduced a ‘coarse-graining’ procedure into the Sample Entropy methodology to verify the structural complexity hidden in high scales, referred to as the Multiscale Sample Entropy (MSE) \cite{Ref23}. This, in turn further spurred the development of MSE algorithms including Composite Multiscale Sample Entropy \cite{Ref62} and Refined Composite Multiscale Sample Entropy \cite{Ref64}. However, due to the ‘coarse-graining’ procedure, the requirement of long data length remains even more problemic and hard to be satisfied in most practical situations. In 2011, Multivariate Multiscale Sample Entropy (MMSE) was introduced which successfully combined dataset from multiple channels to estimate the dynamics of the system more accurately and with shorter data length \cite{Ref24}. The key improvement of MMSE is the form of composite delay vector which involves and reconstructs data segments from multiple channels, whereby the inner correlations among diverse signals are preserved \cite{Ref24}. The existing multivariate multiscale entropies to date include:
\begin{itemize}
    \item Multivariate Multiscale Sample Entropy (MMSE) \cite{Ref24}, a method which performs joint multivariate analysis of physiological signals associated with multiple channels.
    \item Multivariate Multiscale Permutation Entropy (MMPE) \cite{Ref37}, an extension of standard Permutation Entropy \cite{Ref36} which inherits the desirable properties of PE, such as fast computation and simple implementation.
    \item Multivariate Multiscale Fuzzy Entropy (MMFE) \cite{Ref58} which combines Composite Delay Vectors and Fuzzy Entropy \cite{Ref90}, and exhibits more stable smoother estimates than MMSE.
    \item Variational embedding Diversity Entropy (veMDE) \cite{Ref96}, a method developed on the basis of Diversity Entropy \cite{Ref48} that combines angular distance and relative probability, and exhibits a low computational load with similar performance as MMPE.
\end{itemize}

Despite success, the inherent shortcomings of amplitude-based entropy calculation still remain a major obstacle towards their more widespread use . Other issues with current multivariate entropy methods are included as:

\begin{enumerate}
    \item The rule of thumb is that the requirement of data length is around $10^m$ to $30^m$, where $m$ refers to the embedding dimension \cite{Ref71}. Hence, the choice of the embedding dimension is limited by the available sample points.
    \item The ‘coarse-graining’ process further emphasizes the drawback of limited data size, which causes inaccurate and undefined estimation for high scale analysis.
    \item Amplitude-based distance is always sensitive to outliers such as noise and artifacts.
    \item Poor quality of any single channel has a large impact on the multivariate performance.
    \item Excessive computational load is required when implementing multi-channel analyses.
\end{enumerate}

 Recently, Wang \textit{et al}. \cite{Ref96} introduced a new way to combine datasets from multiple channels into one entropy algorithm estimation, termed Variational Embedding Multiscale Diversity Entropy. Here, inspired by \cite{Ref96}, a new multivariate entropy method is proposed based on Sample Entropy named Variational Embedding Multiscale Sample Entropy (veMSE). This new method offers the following advantages as:
 
\begin{enumerate}
    \item Complexity estimates at a higher embedding dimension are better defined with limited data size.
    \item The requirement for the amount sample points is lower than current Sample entropy-based methods.
    \item Strong noise-robustness is exhibited across the scales.
    \item The overall performance of multivariate estimate is independent on the quality of any single-channel within a dataset.
    \item Less computational time is needed owing to straightforward and efficient implementation.
\end{enumerate}

The reminder of the paper is organized as follows:
In Section \uppercase\expandafter{\romannumeral2}, the new veMSE algorithm is outlined. Section \uppercase\expandafter{\romannumeral3} demonstrates the operation of veMSE on simulated signals to give an initial insight with regards to the choice of parameters. Then, based on the suggested parameter setting in Section \uppercase\expandafter{\romannumeral3}, Section \uppercase\expandafter{\romannumeral4} considers and discusses the properties of veMSE, including noise robustness, directionality, and calculation efficiency. Next, veMSE is applied to real-world signals as wind dynamics and heart rate variability in Section \uppercase\expandafter{\romannumeral5}, and compared with the performance of univariate MSE and MMSE. Finally, conclusions are given.

\section{Variational Embedding Multiscale Sample Entropy}
\begin{algorithm}[htb]

    \caption{Variational Embedding Multiscale Sample Entropy}
    \label{veMSE}
    \begin{algorithmic}
    \Statex  Assume that there are $P$ channels measured from a system, where a signal recorded from the $c^{th}$ channel is denoted by $x_c(i)$ and is of length $N$, where $1 \leq c \leq P, 1 \leq i \leq N$. Parameters involved in the veMSE algorithm are the tolerance quotient $(r)$, embedding dimension $(m)$, scale factor $(\tau)$ and time lag $(L)$. The detailed steps of veMSE are shown below.

\begin{enumerate}[ 1)]
    \item Procedure of coarse graining is firstly applied to the original datasets for all the channels. The scaled time series are calculated as
    \begin{equation}
        y^{(\tau)}(j) = \frac{1}{\tau} \sum_{i=(j-1)\tau+1 }^{j\tau}x(i),\quad 1 \leq j \leq N_t,\;N_t = \frac{N}{\tau}.
    \end{equation}
    
    \item For each channel, the embedding dimension is set as a variable. The dimension for the $c^{th}$ channel is calculated as $m(c) = m+c-1$, as listed in Table. \ref{Tab:varied_m}. Therefore, combined with the process of time delay, the embedding delay vector of data $y^{(\tau)}(i),\; (1 \leq i \leq N_t)$ for channel $c$ is designated as template, $Y_c^{(\tau)}(i)$ and calculated as
    \begin{equation}
    Y_c^{(\tau)}(i) = \lbrack y^{(\tau)}(i), y^{(\tau)}(i+L)\; ,\dots, \; y^{(\tau)}(i+n)\rbrack,\; n_c = (m(c)-1)L.
    \end{equation}

    \item Compute the Chebyshev distance between templates $Y_c^{(\tau)}$, where the distance is defined according to the amplitude of the embedding vector as 
   
    \begin{equation}
    \begin{split}
    d = max \lbrace Y_c^{(\tau)}(i+k)) - Y_c^{(\tau)}(j+k) \rbrace, \\
    where \; 0 \leq k \leq m(c)-1,\; 1 \leq i, j \leq N_t - n_c,\; i \neq j
    \end{split}
    \end{equation} 
    
    \item For each channel, the number of segments, $Y_c^{(\tau)}(j)$, within the tolerance level $r$ of $Y_c^{(\tau)}(i)$, is recorded as $B_c(i)$. In other words, $B_c(i)$ is the number of template matched or the number of similar patterns in the dataset based on the boundary set, where the boundary is defined by the tolerance, $r$. Therefore, the local probability of the occurrence of template match for channel $c$ is
    \begin{equation}
    R_c(i) = \frac{B_c(i)}{(N_t - n_c-1)}  
    \end{equation}

    \item Then, the global probability of the occurrence of template match for a channel $c$ is calculated as
    \begin{equation}
    \Phi (c) = \frac{1}{(N_t - n_c)}\sum_{i=1}^{N_t - n_c}{R_c(i)}
    \end{equation}
    
    \item Compute the sum of the global probability for all the channels as 
    \begin{equation}
    \Phi = \sum_{c=1}^P{\Phi(c)}
    \end{equation}
   Recall that unlike the MMSE algorithm (see Appendix.X), the embedding dimension, $m(c)$, is diverse and varied along with the index of channel $c$.

    \item Modify the embedding dimension to $(m(c)+1)$. Hence, $n_c$ is adjusted to $m(c)\times L$ and the Steps 2-6 are repeated to obtain global probability with increased dimension $\Phi(m+1)$.
    
    \item Variational Embedding Multiscale Sample Entropy is finally obtained as
    \begin{equation}
    veMSE = -ln\frac{\Phi(m+1)}{\Phi(m)}.
    \end{equation}
    
\end{enumerate}
    \end{algorithmic}
\end{algorithm}
    
The steps of proposed veMSE is given in Algorithm. \ref{veMSE}. The key improvement of veMSE is that it allows for the varying setting of the embedding dimension for multi-channel signals as illustrated in Table. \ref{Tab:varied_m}, while simultaneously maintaining the information within each channel. Compared to MMSE (see Algorithm. \ref{MMSE} in Appendix), despite the fact that different embedding values can be also achieved for each channel by MMSE, veMSE focuses more on the information between data channels. Indeed, the composite delay vector in MMSE successfully combines the embedding vectors of multichannel signals. However, there is an discrepancy and bias between the similarity among the recombined composite delay vectors and those among embedding vectors of the original datasets. The proposed veMSE estimates the complexity information of signals in multiple channels without influencing the correlation in each individual signal. Secondly, due to the varying embedding dimensions, veMSE yields a weighted contribution to multiple channels. That is, the probability distribution of each channel is diverse, which serve as an amplification of chaotic features in measured signals of multi-channel systems. Thirdly, since the probability of occurrences of similar patterns will be processed multiple times based on the number of channels, and the summation process will be performed before the logarithm operation in the last step, the veMSE is theoretically able to unveil the complexity properties under higher embedding dimension but with the same data length, in comparison with other algorithms based on Sample Entropy.

\begin{table} [htbp]
\small
\caption{Relation between the embedding dimension ($m$) and index of channel ($c$)}
\label{Tab:varied_m}
\begin{center}  
\begin{tabular}{|p{5cm}<{\centering}||p{1cm}<{\centering}|p{1cm}<{\centering}|c|p{1.5cm}<{\centering}|}  
\hline  
Index of channel $(c)$ & 1 & 2 & $\dots$ & $c$\\
\hline 
Embedding dimension $(m(c))$ & $m$ & $m+1$ & $\dots$ & $m+c+1$\\
\hline
\end{tabular}  
\end{center}  
\end{table}

To show the performance of the new proposed veMSE, both synthetic signals and real physical datasets are considered in the following sections. As one of the popular multichannel entropy algorithms, MMSE will be employed as a  benchmark for veMSE under the same conditions.

\section{Results of veMSE on stimulated signals}
In this section, synthetic signals generated based on five models will be utilized to illustrate the performance of veMSE including Gaussian white noise, flicker noise (coloured noise), and autoregressive (AR) models AR$(1)$, AR$(2)$ and AR$(3)$. Standard deviations of all the generated signals were set to $\sigma=1$. Coefficients of AR models are given in Table.\ref{Tab:AR_model}. Parameters that will be discussed in the following subsections include embedding dimension, data length, tolerance, and scale factor. To avoid unknown influence and control variables, time lag were set to $L = 1$ for all operations.

\begin{table} [htbp]
\small
\caption{coefficients of AR models}  
\label{Tab:AR_model}
\begin{center}  
\begin{tabular}{|p{3cm}<{\centering}||p{2cm}<{\centering}|p{2cm}<{\centering}|p{2cm}<{\centering}|}  
\hline  
Coefficients & $a_1$ & $a_2$ & $a_3$ \\
\hline 
AR $(1)$ & $0.5$ & $-$ & $-$\\
\hline
AR $(2)$ & $0.5$ & $0.25$ & $-$\\
\hline
AR $(3)$ & $0.5$ & $0.25$ & $0.125$\\
\hline

\end{tabular}  
\end{center}  
\end{table}

Figures in each subsection are presented in pairs to show the results for five models. Upper panels give the curves of white noise and flick noise, while the panels in the bottom depict the results of AR models in contrast to white noise. Complexity curves of entropy values are plotted as error-bar figures, averaged over outcomes of 20 realizations for each model.

\subsection{Varied Embedding Dimension (m)}

\begin{figure}[htbp]
        \centering 
         \subfigure[]{
        \includegraphics[width=3.4in]{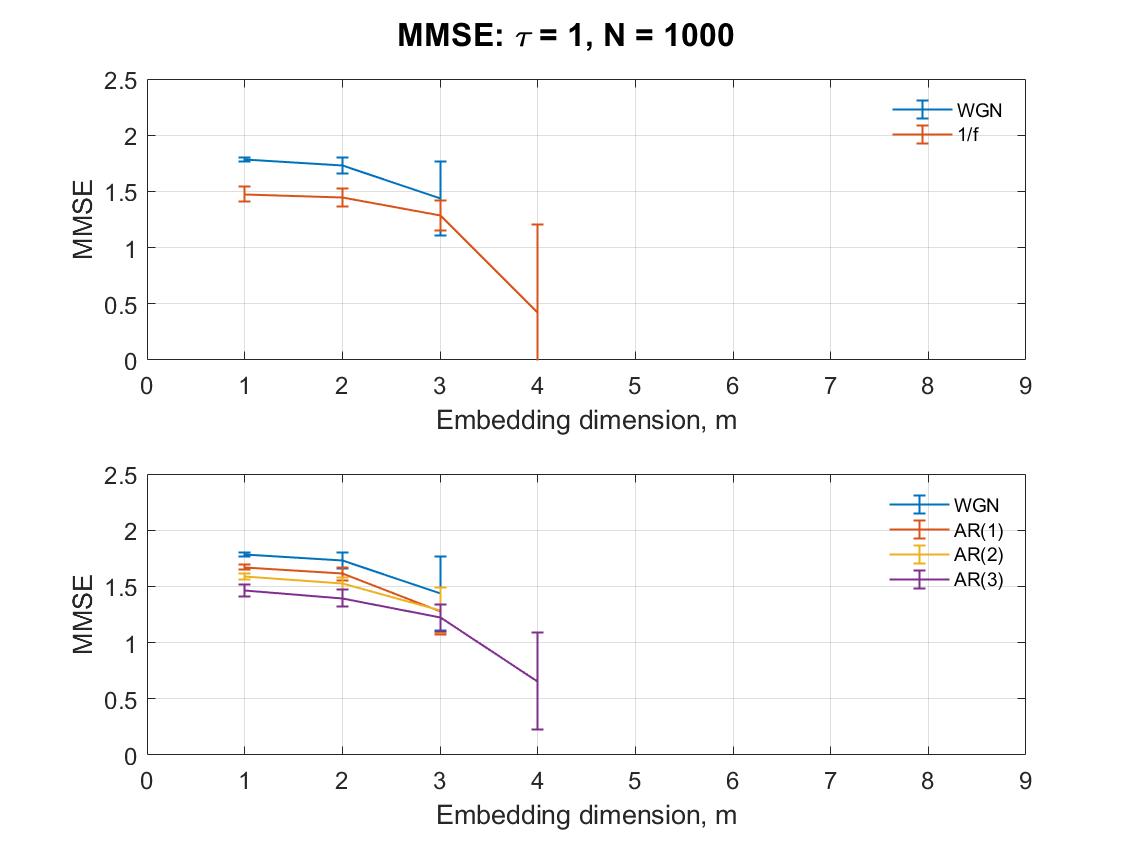}
        \label{fig:vari_m_MMSE}
        }
        \subfigure[]{
        \includegraphics[width=3.4in]{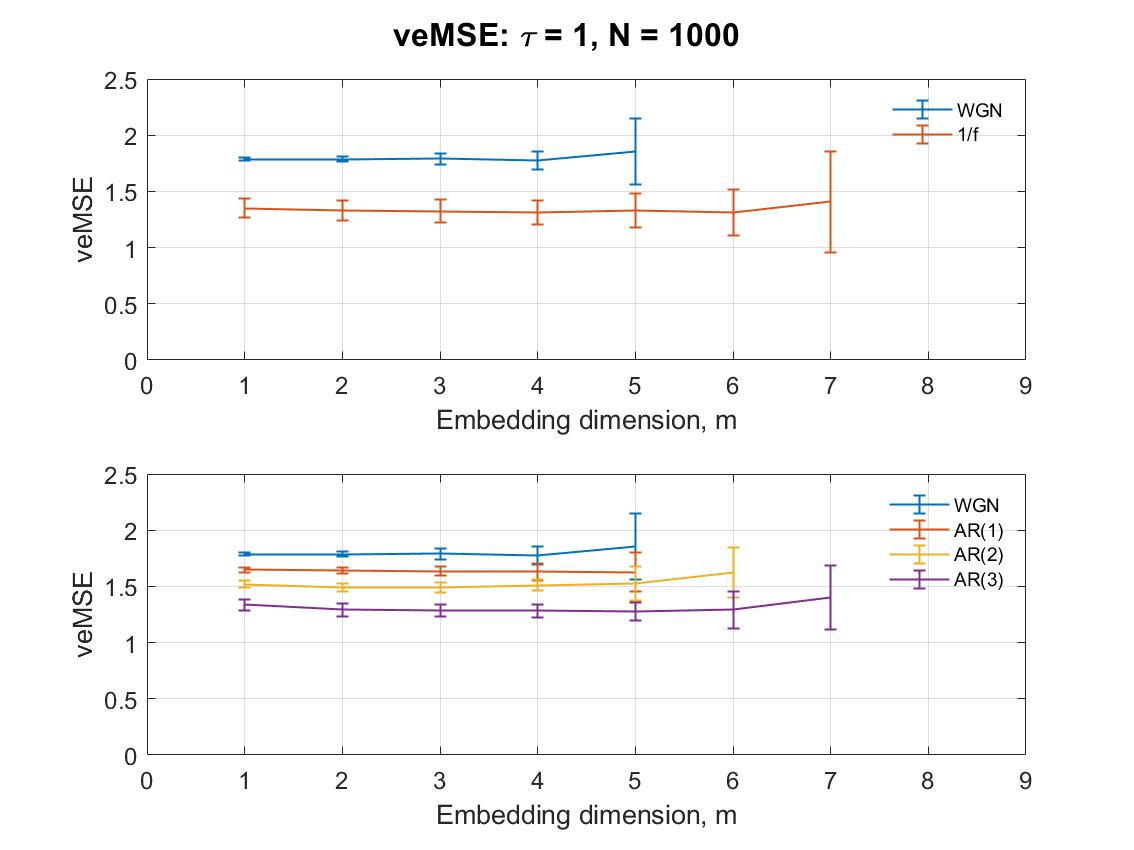}
           \label{fig:vari_m_veMSE}}
        \caption{Operation of single-scale a) MMSE, and b) veMSE as a function of the embedding dimension, m.}
         \label{fig:vari_m}
\end{figure}

Usually, for the implementation of SampEn-based algorithms, the embedding dimension and data length are two parameters that are interdependent and mutually coupled. Data size is restricted to between $(10^m-30^m)$ as a rule of thumb \cite{Ref82}. In real world, signals recorded do not have infinite length and are generally limited by operation time and memory space. Therefore, the embedding dimension is commonly set to $m=2$ or $m=3$ for a signal with 1000 samples \cite{Ref26}. Higher values of embedding dimension with a shortage of data size will cause unstable estimation as standard MMSE given in Figure.\ref{fig:vari_m_MMSE}.

Results of veMSE as a function of embedding dimension are shown in Figure.\ref{fig:vari_m_veMSE}. Each entropy is calculated based on signals from two channels. Except the independent variable $m$, other parameters, such as scale factor and data length, are set as constant values (1 and 1000 respectively). The tolerance, $r$ is varying according to the total variance of the covariance matrix of processed data sets as $r\times tr(S)$, and here the tolerance quotient was fixed to $r = 0.15$.

Figure.\ref{fig:vari_m_veMSE}, where embedding dimension is modified from 1 to 9, shows that unlike MMSE, signals with a complex correlated structure are able to give a defined entropy value even at high embedding dimensions, as e.g. AR(3) in scale of 7. Also, signals with higher randomness are more likely to become unstable as the embedding dimension increase. However, even in the case of Gaussian white noise with highest randomness, here veMSE with the embedding dimension equal to 5 was able to successfully and stably process the data. On the other hand, traditional Multiscale Sample Entropy methods fail to give a defined value with the embedding dimension higher than 3 under the same condition \cite{Ref1}. Therefore, from the viewpoint of estimation stability, veMSE exhibits a marked improvement when it comes to complex information in high dimensions. 

\subsection{Varied Data Length (N)}
The data length of the signal is another limitation in addition to embedding dimension when implementing entropy-based calculation, particularly in real world processes. Amplitude-based entropy algorithms require at least 1000 data points to guarantee a consistent estimation such as with Multiscale Sample Entropy (MSE) and Multiscale Fuzzy Entropy (MFE) \cite{Ref90}. However, in real world data, as in the analysis of heart rate variability for example, to obtain the required data size for R-R intervals, a minimum of 5 minutes of the raw Electrocardiograph (ECG) signals are needed. In practice, the implementation of such a long-time recording in a controlled state is hard to be satisfied. Compared to amplitude-based entropy methods, space distance-based entropy algorithms such as Cosine Similarity Entropy, show less restriction to data length, with a minimum of 700 samples required \cite{Ref1}.

\begin{figure}[htbp]
        \centering 
         \subfigure[]{
        \includegraphics[width=3.4in]{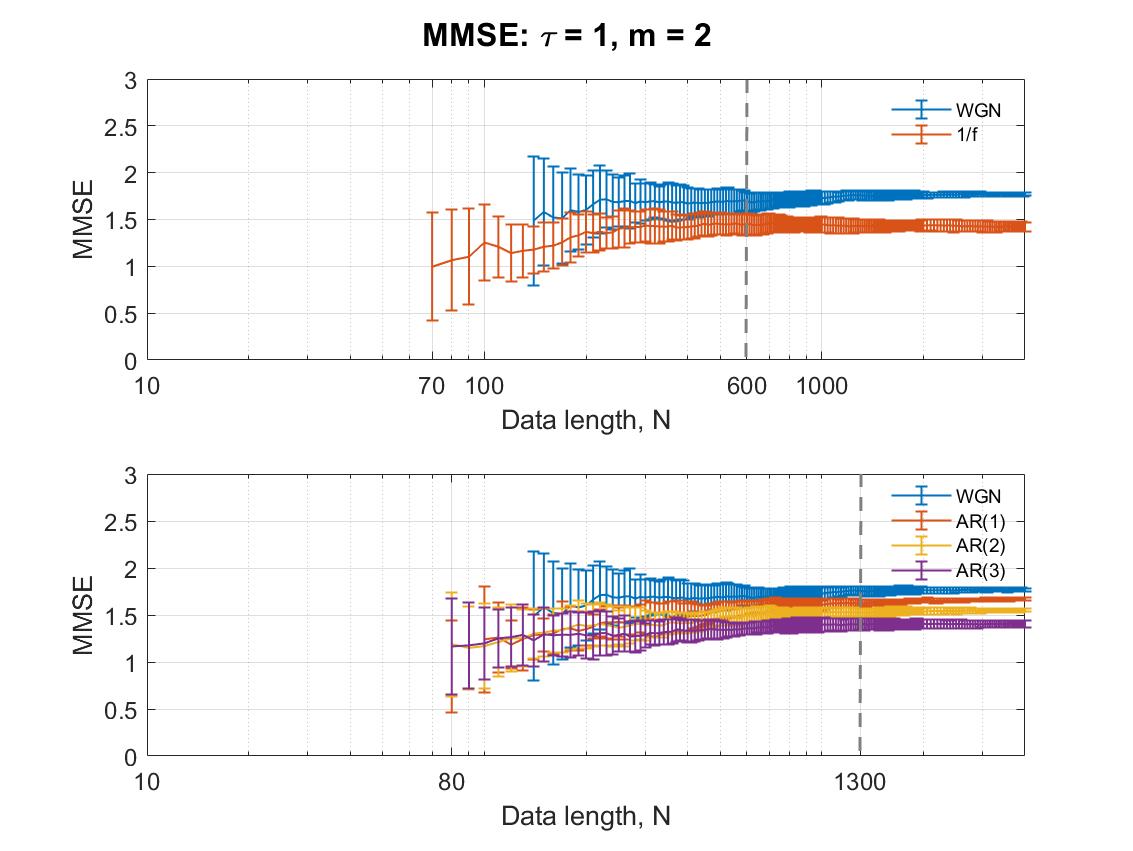}
        \label{fig:vari_N_MMSE}
        }
        \subfigure[]{
        \includegraphics[width=3.4in]{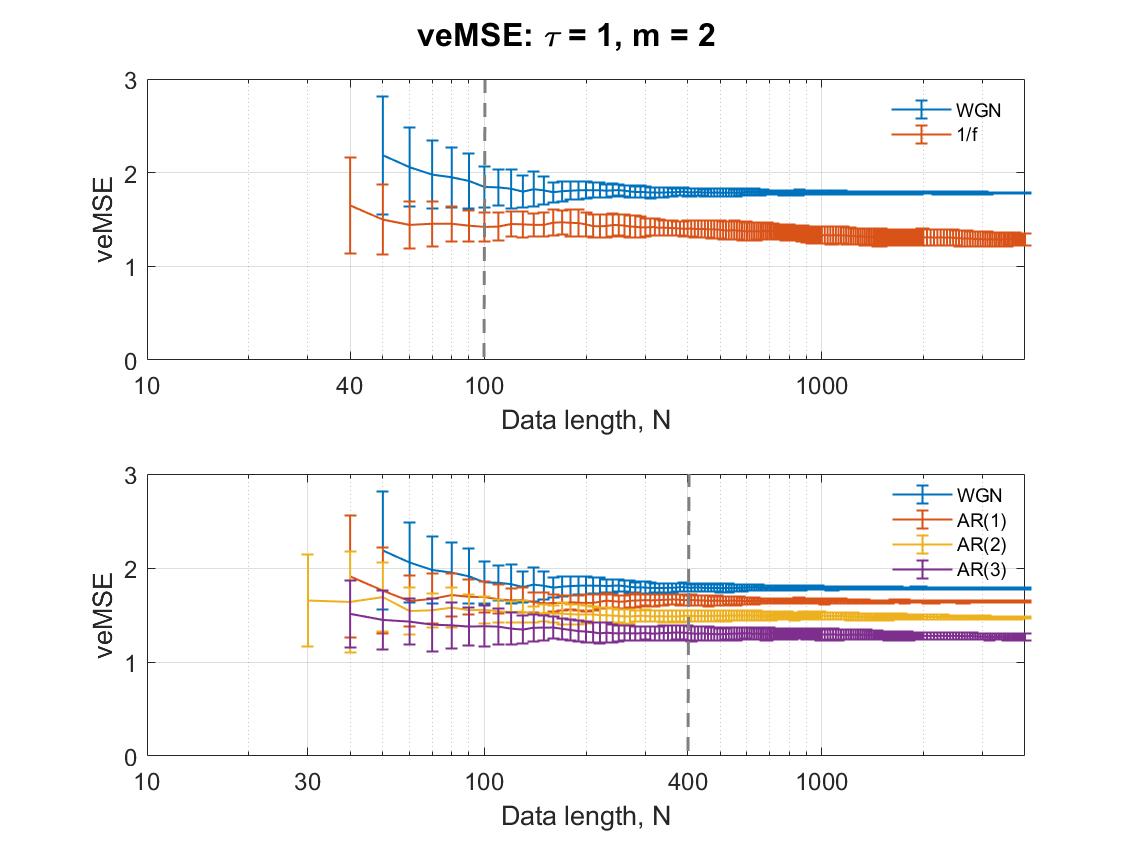}
           \label{fig:vari_N_veMSE}}
        \caption{Operation of single-scale a) MMSE, and b) veMSE as a function of the data length, N.}
         \label{fig:vari_N}
\end{figure}

Figure.\ref{fig:vari_N_veMSE} illustrates the performance of single scale of veMSE as a function of data length in a logarithmic scale. The embedding dimension was set to $m=2$ and the choice of tolerance was the same as before. The values of veMSE for white noise and $1/f$ noise were not defined before $N$  = 40, while AR(2) was not defined with veMSE when $N$ was smaller than 30. The smallest sample sizes for AR(1) and AR(3) were also $N$ = 40.

The standard deviation of the entropy results is gradually narrowing down with an increase in data length, while the range for each error-bar is ascending from top to bottom. As a result, a system with higher structure (AR(3)) reveals larger standard deviation. In addition, the consistency of the estimation can be guaranteed as evidenced by the relative position of curves in each graph maintains being unchanged as data length, $N$, increases. More importantly, when analysing the white noise and flicker noise in the top graph, the estimation at $N$ = 100-sample length could successfully separate the complexity degrees of the two signals, while in the bottom graph, when the data length reaches $N$ = 400 samples, there is no intersection region among entropy values from different models. This illustrates that the requirement of data length when applying veMSE is much lower than other entropy methods, e.g. MMSE in Figure.\ref{fig:vari_N_MMSE} gives the $N$ = 1300 is required for the separation of 4 models. Subsequently, this property gives the way to reveal the complexity information under high scales, with limited data length, but with a stable estimation, which better serves the balance between the dimension and data size.

\subsection{Varied tolerance (r)}

\begin{figure}[htbp]
        \centering 
         \subfigure[]{
        \includegraphics[width=3.4in]{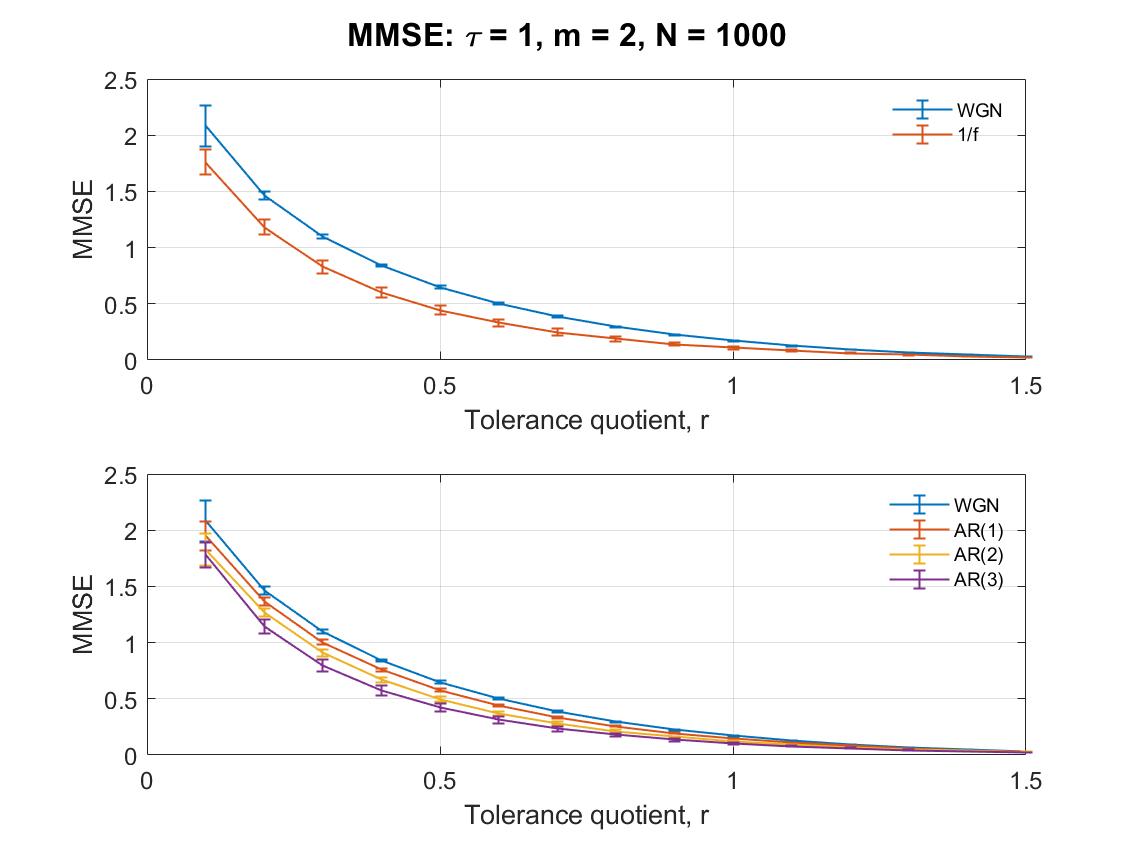}
        \label{fig:vari_r_MMSE}
        }
        \subfigure[]{
        \includegraphics[width=3.4in]{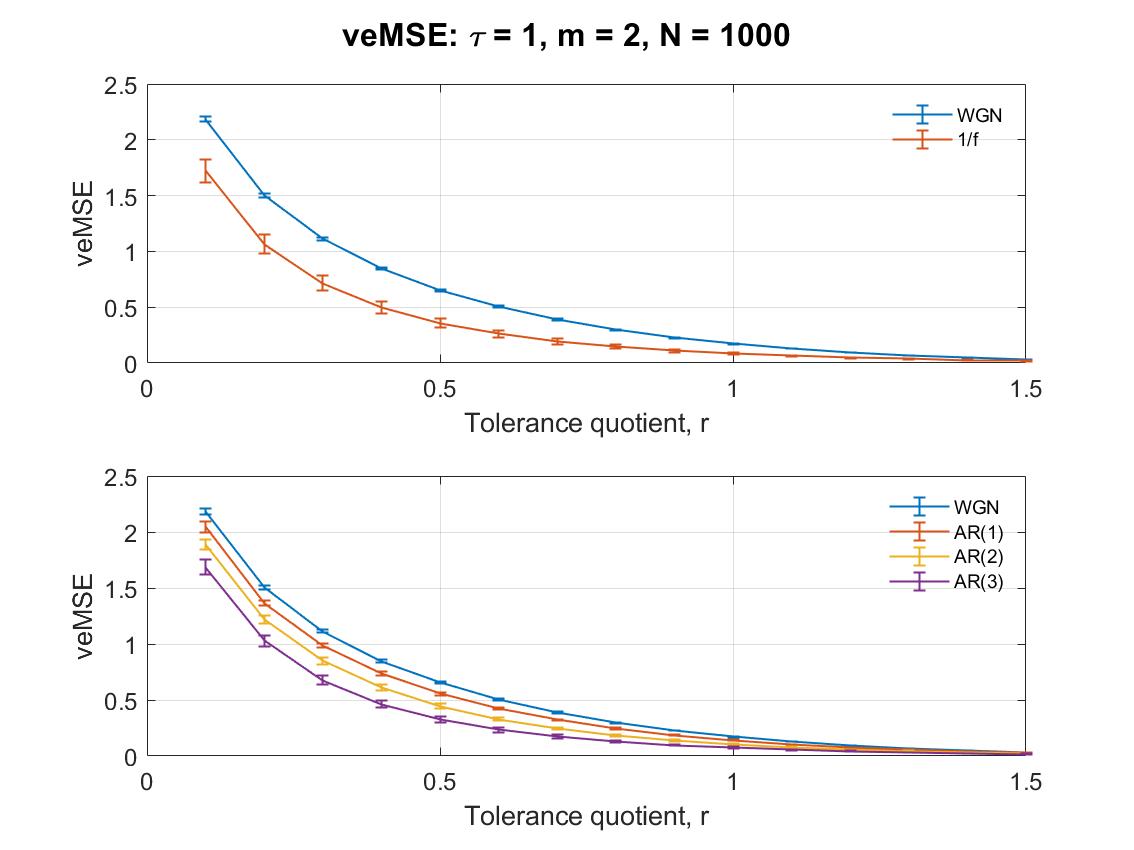}
           \label{fig:vari_r_veMSE}}
        \caption{Operation of single-scale a) MMSE, and b) veMSE as a function of the tolerance, r.}
         \label{fig:vari_r}
\end{figure}

The tolerance, $r$, can be explained as the boundary of the similarity degree among comparing templates. The SampEn-based algorithms limit the tolerance to a hard threshold, as the Heaviside function in related to the standard deviation of the original data. However, for a multivariate case with multichannel data sets, only a single value is needed in the algorithm. As in \cite{Ref24}, the choice of tolerance of veMSE is dependent on the total variance of the covariance matrix, \textbf{$S$}, of the analysed data sets. Therefore, the tolerance was set as $r\times tr(S)$. 

Figure.\ref{fig:vari_r_veMSE} illustrates single scale entropy estimation as a function of tolerance quotient, $r$, varying from 0.1 to 1.5 at 0.1 increments. The data length and embedding dimension were separately fixed at $N$ =  1000 and $m$ = 2 to show the influence of varied tolerance setting. Observe from the figures that for all curves, the increase of tolerance quotient results in a monotone decrease in complexity estimation, which is the same as MMSE in Figure.\ref{fig:vari_r_MMSE}. All can be obviously distinguished from each other before $r $ = 1 in veMSE. The values after $r$ = 1 are too small to be differentiated. The gap among different complexity estimations in the two figures obviously narrows down after r = 0.5, therefore, supporting that the value of tolerance quotient to be chosen below $r$ = 0.5.

\subsection{Varied Scale Factor}

As noted by Costa \textit{et al}. in \cite{Ref23}, the multiscale analysis by integrating consecutive coarse-graining procedure is of importance in signal processing associated with hidden correlation structure in data. Based on the aforementioned analysis of parameters involved in veMSE, the embedding dimension was set to $m$ = 2 and the tolerance was chosen to $r$ = 0.15 multiplied by total variance of the covariance matrix. With regard to performance of multiscale analysis, graphs of multichannel entropy results are presented in response to the scale factor varying from $(\tau)$ = 1 to $(\tau)$ = 40. Dual channel data with $N$ = 3000 sample data points for each model were considered.

\begin{figure}[htbp]
\centering
\begin{minipage}[t]{3.3in}
\centering
\includegraphics[width=3.1in]{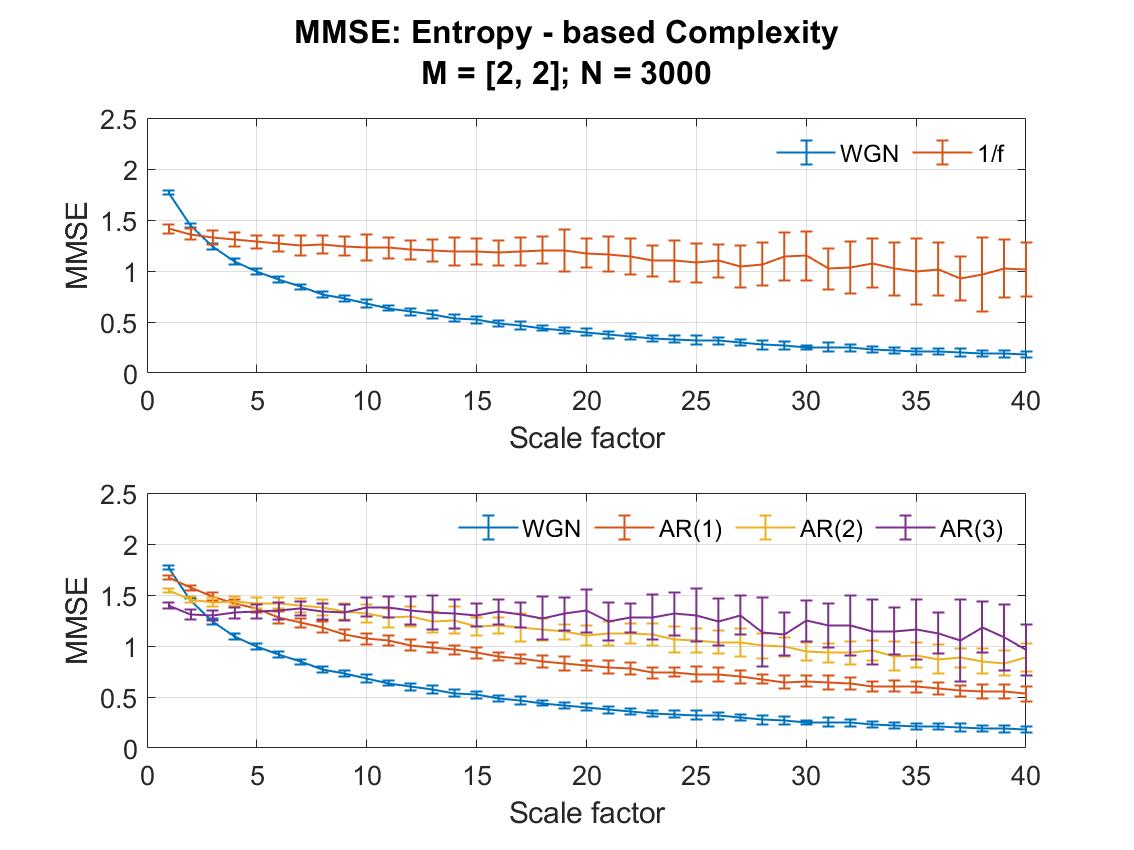}
\caption{Operation of Multivariate Multiscale Sample Entropy \cite{Ref24} with the embedding dimension set as [2 2].}
\label{fig:MMSE_22}
\end{minipage}
\hspace{0.1in}
\begin{minipage}[t]{3.3in}
\centering
\includegraphics[width=3.1in]{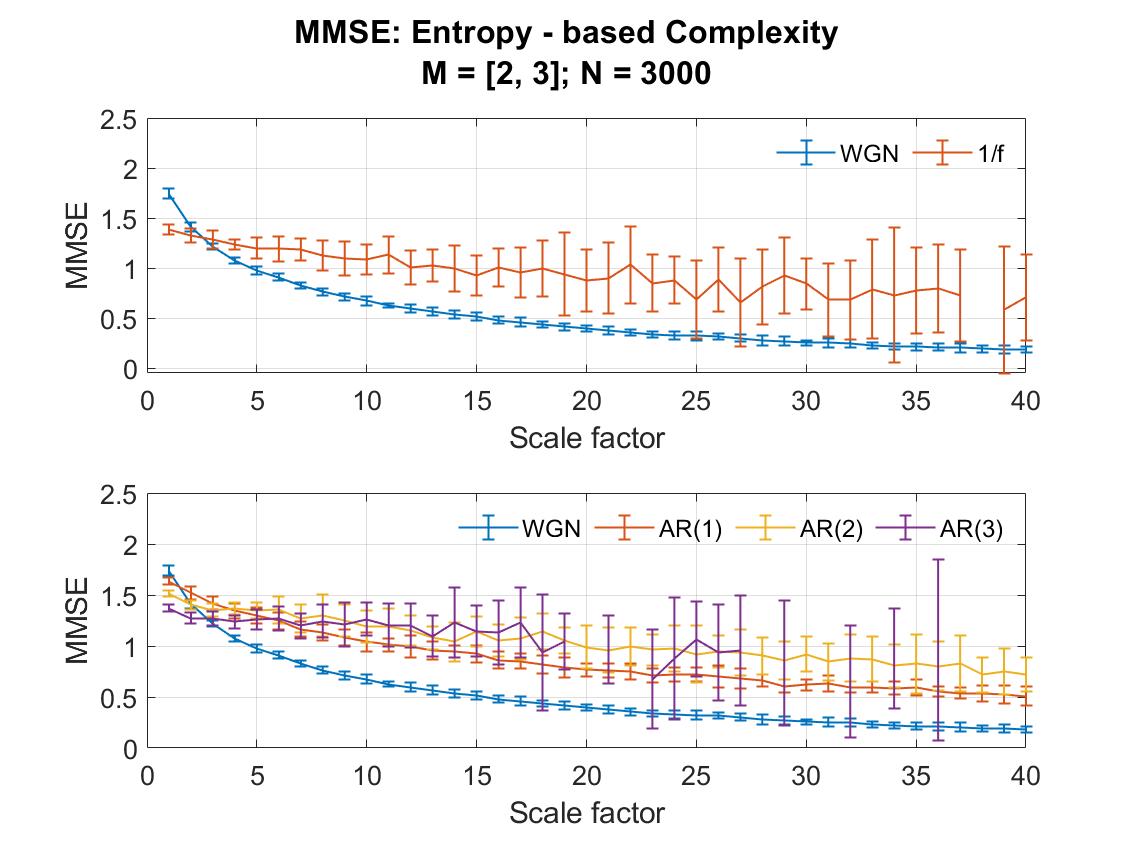}
\caption{Operation of Multivariate Multiscale Sample Entropy \cite{Ref24} with the embedding dimension set as [2 3].}
\label{fig:MMSE_23}
\end{minipage}
\end{figure}

\begin{figure}[htbp]
\centering
\begin{minipage}[t]{3.3in}
\centering
\includegraphics[width=3.1in]{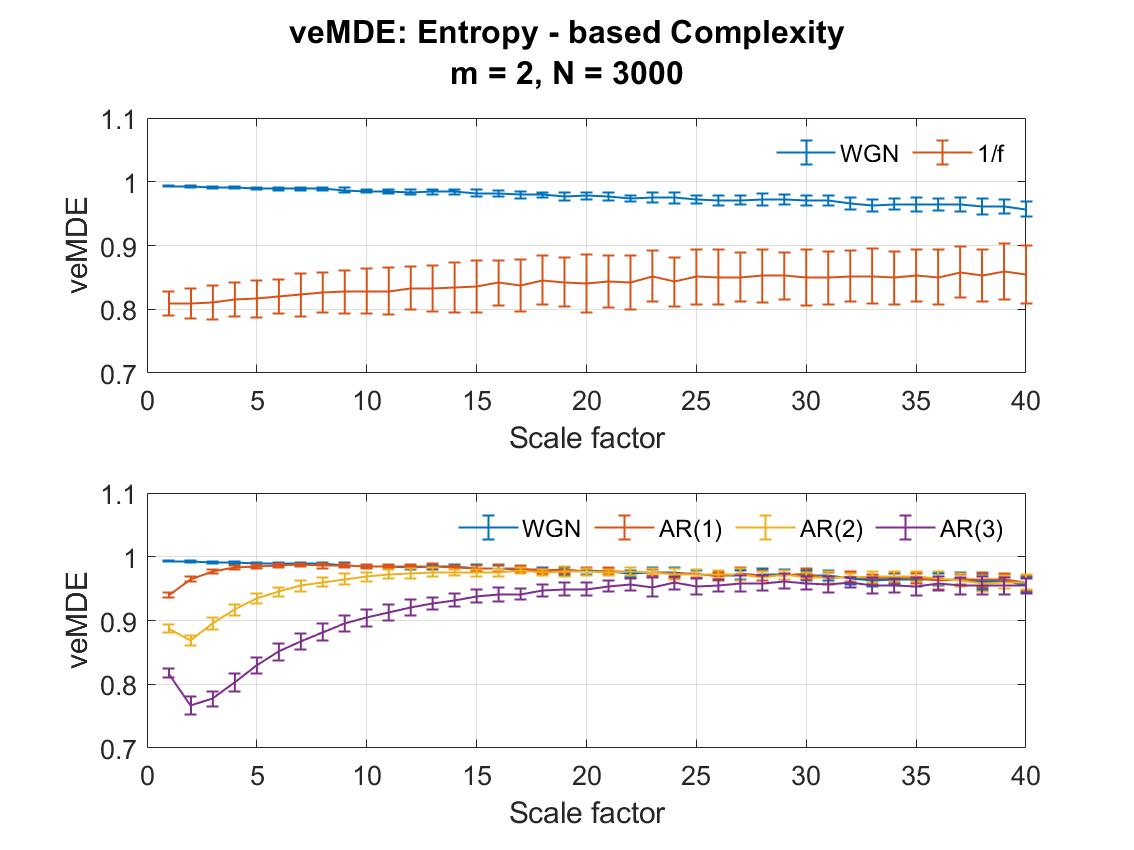}
\caption{Operation of variational embedding Multiscale Diversity Entropy  \cite{Ref96} with the embedding dimension set as 2.}
\label{fig:veMDE}
\end{minipage}
\hspace{0.1in}
\begin{minipage}[t]{3.3in}
\centering
\includegraphics[width=3.1in]{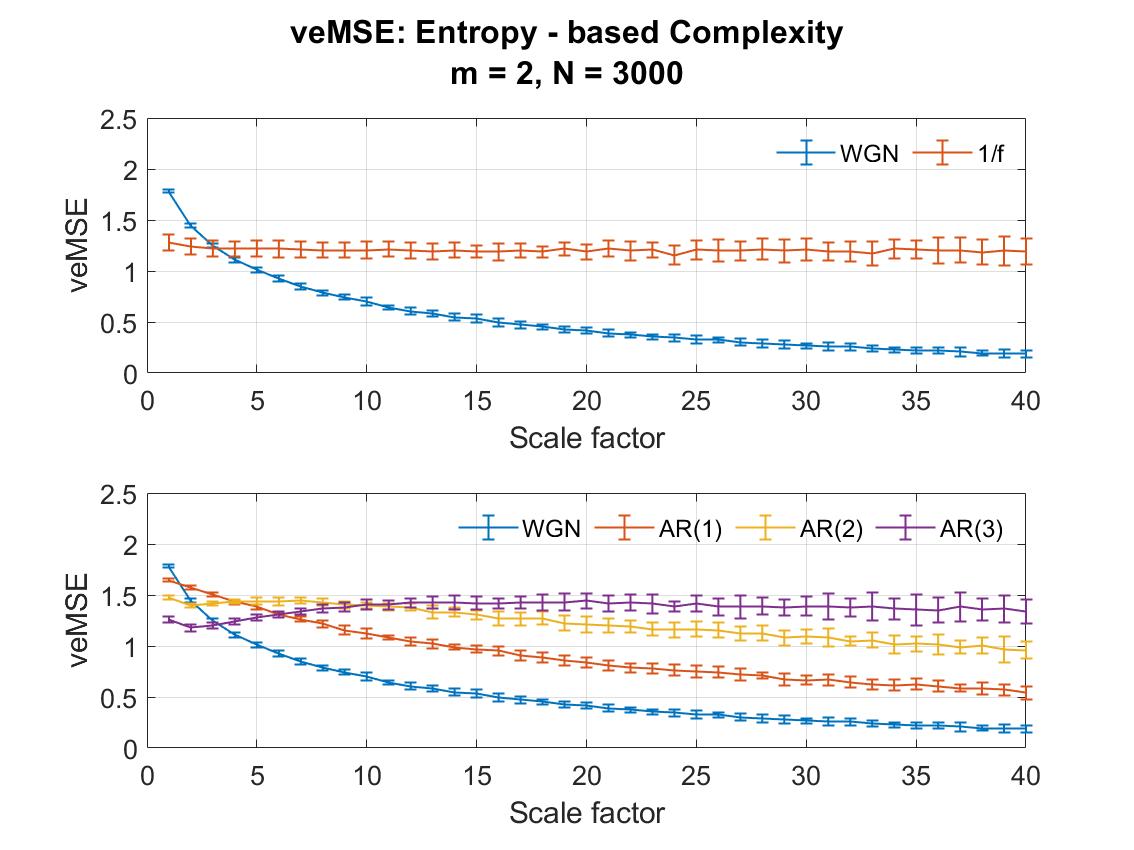}
\caption{Operation of the proposed variational embedding Multiscale Sample Entropy with the embedding dimension set as 2.}
\label{fig:veMSE}
\end{minipage}
\end{figure}

To further elucidate the extent of improvements of veMSE over the Multivariate Multiscale Sample Entropy (MMSE) and Variational Embedding Multiscale Diversity Entropy (veMDE), their performances were compared with the proposed veMSE method. With the same data size, due to the variational dimension feature of veMSE, two different settings of the parameter related to embedding dimension for MMSE were applied and are shown in Figure.\ref{fig:MMSE_22} and Figure.\ref{fig:MMSE_23}. The performance of veMDE is given in Figure.\ref{fig:veMDE}, with the embedding dimension set to $m$ = 2, while the results of veMSE are presented in Figure.\ref{fig:veMSE}.

These figures demonstrate that complex structure hidden in higher dimension is hard to be unveiled by MMSE. The standard deviation of those 20 realizations grows steadily as the scale factor increases. Overall, the dimension setting pair $[2,2]$ gives a better performance in MMSE for the considered restricted data length. However, even under the optimal dimension settings, as in Figure.\ref{fig:MMSE_22}, the complexity of AR(3) in purple and AR(2) in yellow fails to be distinguished in multi-scale cases. As for veMDE in Figure.\ref{fig:veMDE}, Diversity Entropy is developed based on angular distance and distribution probability \cite{Ref48}. As can be seen in the graph, veMDE reveals a consistent estimation for each system, which exhibites a short-term correlation. Nevertheless, long-term structural complexity of each model fails to be shown by veMDE. Besides, due to the lack of ability to estimate long-term correlation, veMDE for AR models which contain highly correlated structure can not be distinguished from white noise in large scales.

On the other hand, consistent with the advantages as aforementioned theoretical discussion, the merits of veMSE can be clearly seen from in Figure.\ref{fig:veMSE}. To better specify the improvement, the optimal dimension setting [2 2] of MMSE in Figure.\ref{fig:MMSE_22} was utilized to compare with veMSE. Observe from the upper graphs of the two types of noise, although both of the two algorithms were able to distinguish between the two models, the complexity of white noise went down while the flicker noise maintained a certain level in spite of the increasing scale. The range of entropy in error-bars for flicker noise based on veMSE was much narrower than that in basis of MMSE especially in large scales. Secondly, in the bottom graph, values of AR(2) and AR(3) (in yellow and purple) fail to be fully separated by MMSE in the cases of high scale, as stated above, while with the same data length, the separability of AR models with different orders can be successfully accomplished in veMSE. It is critical to apply the entropy calculation under multiscale situation since the long-range correlation of the system is largely ignored in the analysis under low scale. Next, it can be observed that minor difference exists between the complexity estimation of the two models, referred to white noise and AR(1) (blue and red line in bottom graph). Instead, the enhancement produced by veMSE is particularly revealed on the analysis of highly correlated and structural signals, systems with higher structural complexity.

Overall, according to the comparison of veMSE and MMSE as well as veMDE based on the above five models, the veMSE provides a more stable estimation that can better demonstrate complex temporal fluctuations. In addition, veMSE is especially suitable for multiscale analysis of highly correlated signals which exhibit variation of spatial-temporal patterns over a range of scales.

\section{Properties of veMSE}

We now elaborate on the three properties of the proposed entropy veMSE algorithm: noise robustness, directionality, and calculation efficiency. The parameters setting: data size $N$ = 3000; embedding dimension $ m$ = 2; tolerance $ r$ = 0.15; and scale factor $\tau$ = 1. A bivariate system was considered in the analysis. The results which depict the noise analysis and directionality analysis based on proposed veMSE are presented in Figure.\ref{fig:veMSE_noise} and Figure.\ref{fig:veMSE_dire} respectively. And corresponding performance of MMSE are exhibited in Figure.\ref{fig:MMSE_noise} and \ref{fig:MMSE_dire}. The curves of time requirements of veMSE and MMSE are given in Figure.\ref{fig:time}.

\subsection{Noise Robustness}

\begin{figure}[htbp]
        \centering 
         \subfigure[]{
        \includegraphics[width=3.4in]{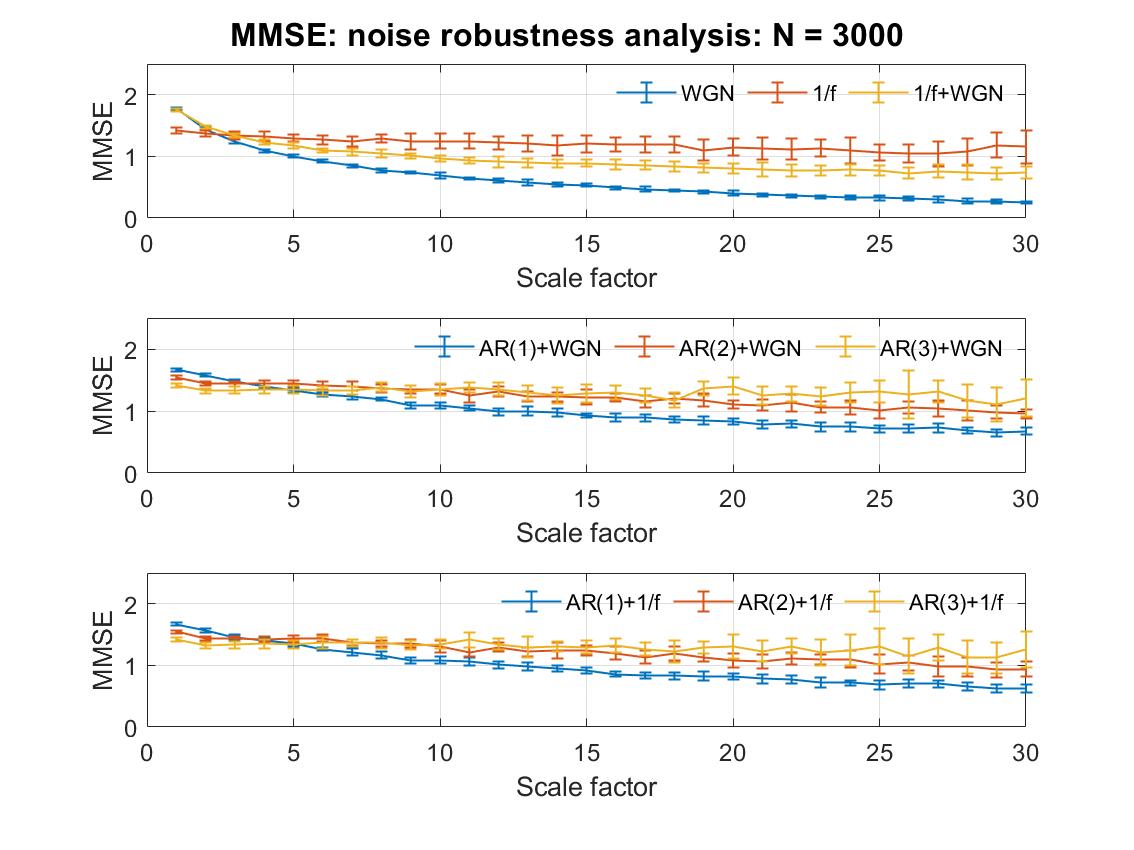}
        \label{fig:MMSE_noise}
        }
        \subfigure[]{
        \includegraphics[width=3.4in]{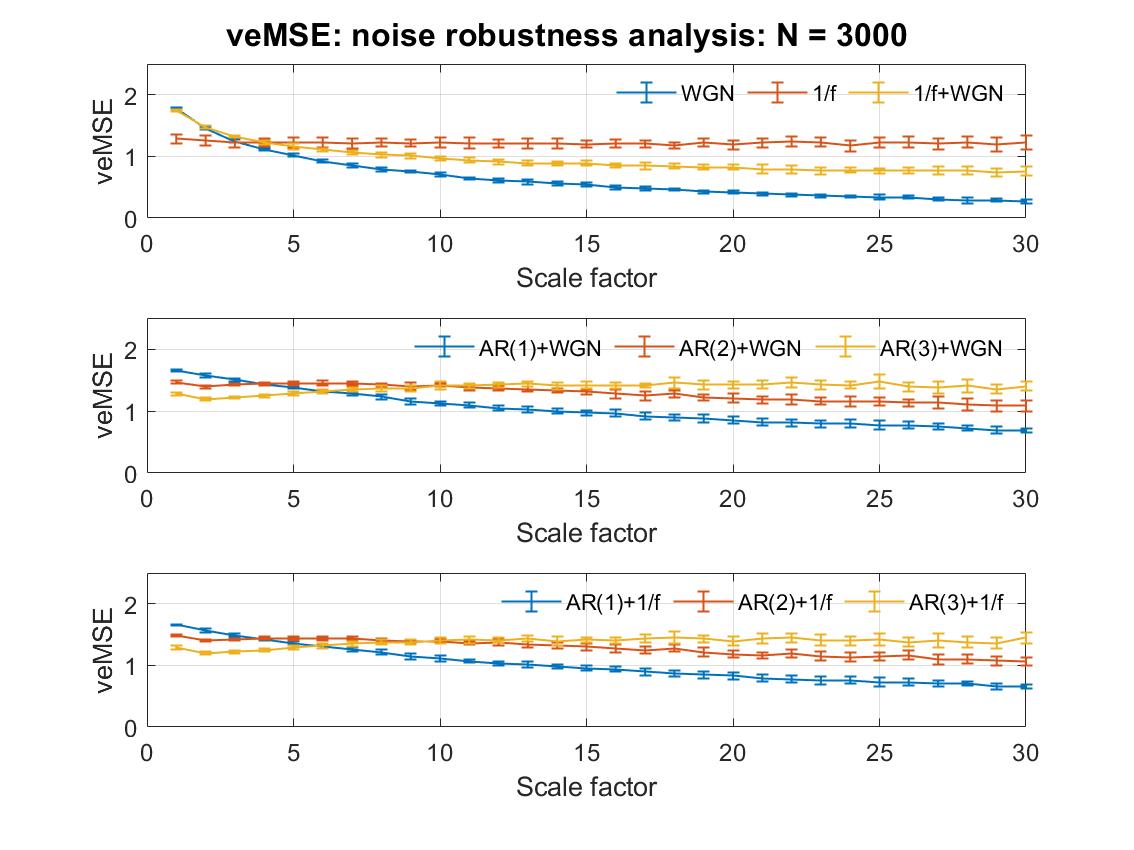}
           \label{fig:veMSE_noise}}
        \caption{Illustration of noise robustness of a) standard MMSE and b) the proposed veMSE.}
         \label{fig:noise}
\end{figure}

Robustness towards noise and artifacts is of critical importance in any estimation. Given that it is infeasible to avoid the noise associated to recording equipment and the ubiquity of artifacts in bio signals, for instance, muscle and electro-magnetic artifacts pervasively exist in EEG-based monitor \cite{Ref137}, the noise-robustness property was tested by comparing complexity estimation for AR models with and without noise. In Figure.\ref{fig:veMSE_noise}, from top to the bottom, the first panel presents the curves for uncorrelated white Gaussian noise (WGN), correlated flicker noise ($1/f$ noise) and coloured noise containing both WGN and $1/f$ noise. It is clear that three systems with different degrees of correlation can be separated by the veMSE. Adding white noise will enhence the short-term correlation as shown at the beginning in the top graph, where the yellow line (\texttt{1/f + WGN}) is as high as blue line (\texttt{WGN}), while the long-term correlation is lower as scale factor increases (\texttt{1/f} $>$ \texttt{1/f}+\texttt{WGN} $>$ \texttt{WGN}). The first graph reveals that veMSE could correctly give a complexity estimation, in line with the theoretical analysis, on the basis of uncorrelated and correlated noise.

In the second and third panel of Figure.\ref{fig:veMSE_noise}, the results of veMSE for AR models with uncorrelated white noise and correlated flicker noise are presented, to contrast to the outcomes of pure AR signals in Figure.\ref{fig:veMSE}. The amplitude of the added noise signal was set to $20\%$ that for the AR signals. Compared to Figure.\ref{fig:veMSE}, the gaps between the complexity curves for the AR models of varying order decrease with noise. However, even that the gap among distinct models is narrowed down, separation can still be achieved to high scales in $20\%$ noisy scenarios. While in case of MMSE, noisy AR signals with different complexity cannot be well divided and the impact of noise was clearly shown in Figure.\ref{fig:MMSE_noise}.  Given these points, the performance of complexity estimation based on veMSE is consistent with cases without noise. A unique feature of veMSE not present in the other MSE algorithms, showing its potential to be applied in practical recording data sets.

\subsection{Directionality}

\begin{figure}[htbp]
        \centering 
         \subfigure[]{
        \includegraphics[width=3.4in]{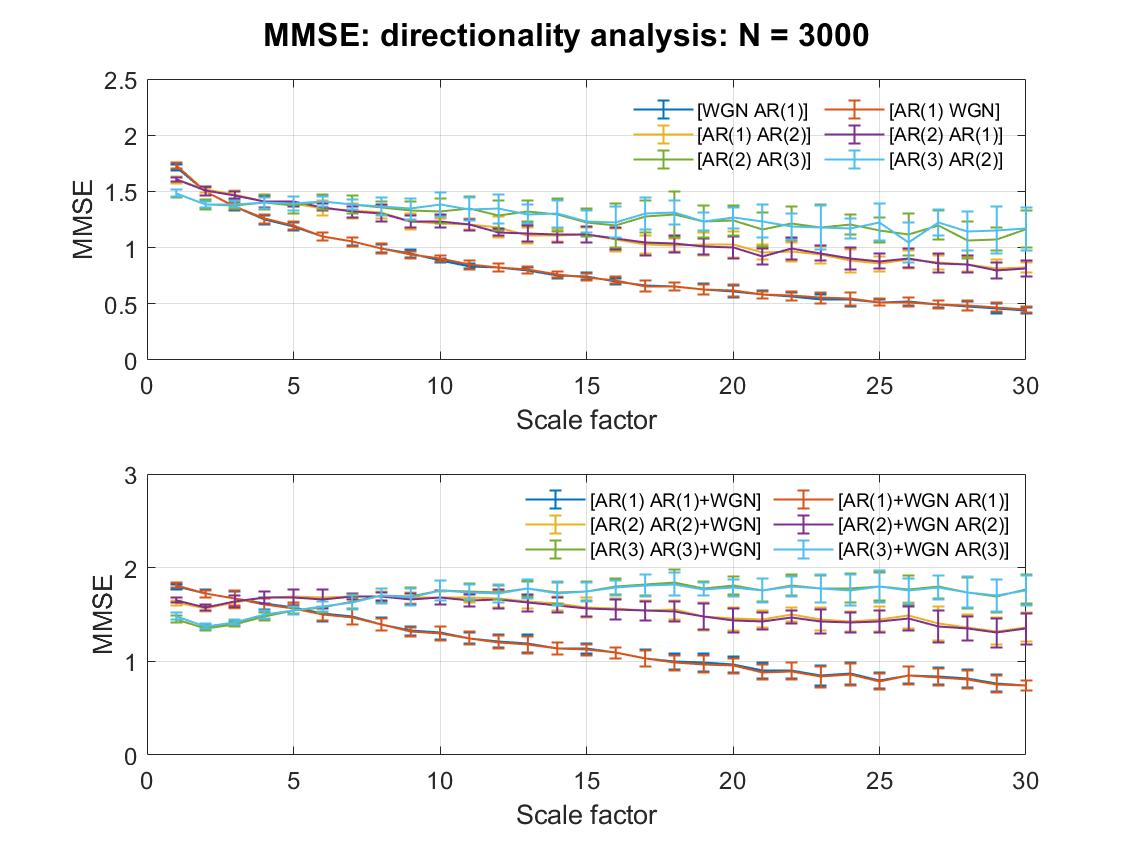}
        \label{fig:MMSE_dire}
        }
        \subfigure[]{
        \includegraphics[width=3.4in]{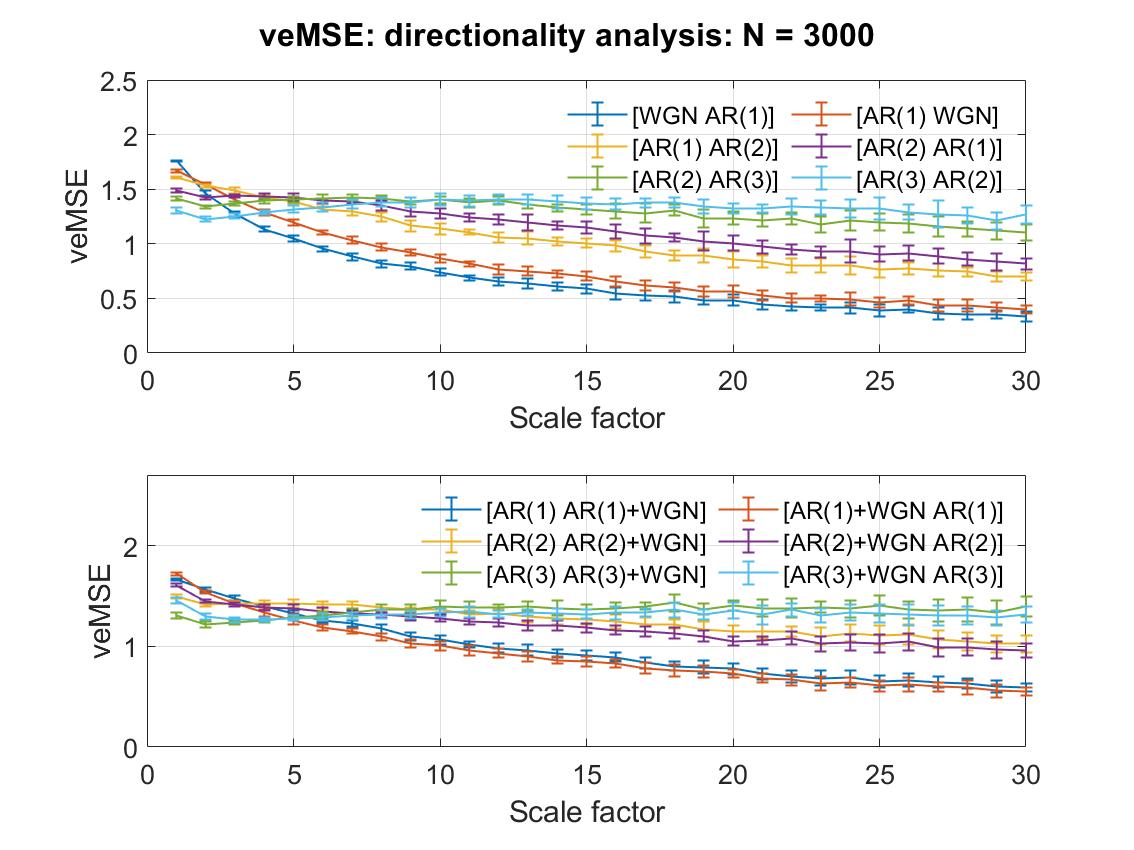}
           \label{fig:veMSE_dire}}
        \caption{Illustration of directionality of a) standard MMSE and b) the proposed veMSE.}
         \label{fig:dire}
\end{figure}

Next feature of veMSE that is going to be discussed is directionality. For multivariate analysis, a problem is that the optimal ordering of the input channel is unknown. Yet, without prior knowledge related to the optimal channel order, the performance of the estimation will be impacted. To this end, the directionality of the veMSE is analysed for bivariate systems. Figure.\ref{fig:veMSE_dire} reveals two graphs, each contains three pairs of curves. In top graph, there are white noise with AR(1), AR(1) with AR(2) and AR(2) with AR(3), with the order of input shown by legend (first present, first processed). As can be seen from this figure, the lower scale estimate is mainly influenced by the first input signal. For example, the blue line \texttt{[WGN AR(1)]} can be clearly recognized lower than the red one \texttt{[AR(1) WGN]} at the beginning especially, in the single scale case. As scale increases, the two lines approach the same level. Similar trend can be seen for the other two pairs.

In the second graph of Figure.\ref{fig:veMSE_dire}, the analysed signals are AR(1), AR(2) and AR(3) with one of the signals in each system associated with white noise. The legend \texttt{[AR AR+WGN]} refers to cases where noise-free signals are the first variate followed by noisy signals and vice versa for \texttt{[AR+WGN AR]}. This setting of the inputs was used to stimulate real scenarios when dealing with multi-channel signals, whereby one of them represents a poor recording with noise. Considering the noise-robustness property of veMSE, the amplitude of noise signals in this subsection is enlarged to the same level as for AR signals, to demonstrate a clear difference when the input order is modified. As shown in the figure, the inverted input orders can be reflected by different start levels, while the curves then approach each other as well as end with similar estimates. Therefore, regardless of the input order, the separation of complexity levels of AR models can be successfully achieved. And observed in Figure.\ref{fig:MMSE_dire} in case of MMSE, the inverted input exhibited no influence on the resulted curves in small scales, while similar performance as veMSE in the larger scale.

As demonstrated in Figure.\ref{fig:veMSE_dire}, the reversed order has limited influence on the estimation in high scales as all the paired curves approach to the same three regions, respectively, while in spite of the modified order, the three models can be separated. However, as similar phenomenon is shown in the top graph, the varying order of the input signal will generate entropy values with different degrees in small scale analysis when the input signals contain distinct structure. Therefore, the direction of the input order needs to be carefully considered when applying small scale analysis, and such consideration can be ignored in high scale analysis with identical system measurement.

\subsection{Computational Complexity}

Admittedly, entropy analysis based on multichannel signals is more time-consuming than other single variate estimation as it should be. Nevertheless, for potential applications as real time processing in the future, calculation efficiency is one of the critical factors that need to be carefully considered. Therefore, in this subsection, time consumption of veMSE is discussed and compared with commonly used MMSE and the results are given below.

\begin{figure}[htp]
    \centering
    \includegraphics[width=7in]{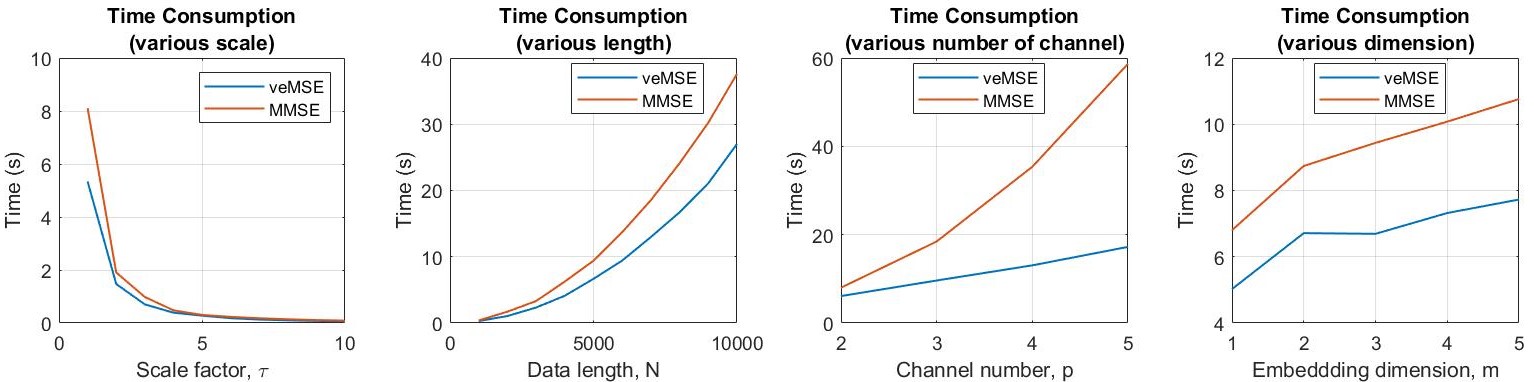}
    \caption{Computation time for MMSE and the proposed veMSE with modified parameters for white Gaussian input.}
    \label{fig:time}
\end{figure}

Figure.\ref{fig:time} shows the processing time as a function of various modified parameters when implementing the veMSE (blue line) and MMSE (red line). All the curves are given as an average over 10 implement realizations. Each graph is designed to reflect the behaviours for only one modified parameter, when the independent variables in the following figures, from the left- to the right-hand side, as the scale factor, length of data, number of channels and embedding dimension. All the entropy calculations were set as single-variate and bivariate processing by default. The data length and embedding dimension were irrelevant variables were fixed to $N$ = 5000 and $m$ = 2.

Overall, the red line, representing computational time of MMSE, is above the blue, that of veMSE, for all the scenarios. Actually, the increase of scale factor and the decrease of data length reflect that when the data size after ‘coarse-graining’ procedure is lower than 1000, the times needed for the two calculations are similar, as shown in the first graph where the scale factor is higher than 5 or when the data length is shorter than 1000 in the second graph. As for the influence of channel number, demonstrated in the third graph, it is reasonable that MMSE need more time as the channel number increases, because the key step for Sample Entropy is the ratio of conditional probability for similar patterns between the embedding dimension, $m$, and its increase, $m+1$, the number of possible ways to apply the  ($m+1$) dimension is equal to the number of channels involved when forming the composite delay vector in MMSE. Therefore, the calculation with an increased embedding dimension will be repeated $c$ times in MMSE, where $c$ denotes the number of channels. Finally, in the relationship between an increased dimension and computation time, the time difference roughly maintains a fixed value in spite of the dimension changing in the last graph.

In general, with reference to the widely applied MMSE, the time needed for the same amount of data with veMSE is shorter. This reflects that the calculations efficiency of veMSE is higher than MMSE, which is of high potential interest to implement real-time monitoring of human states in the future.

\section{Performance of veMSE on real signals}
\subsection{Wind Dynamics}
Having illustrated how veMSE performs on synthetic signals, we now examine the performance of veMSE in real-world systems. In the first place, wind dynamics were examined. The long-term correlations in wind dynamics has been revealed in previous studies by using detrended fluctuation analysis (DFA) \cite{Ref275} and standard Multivariate Multiscale Sample Entropy (MMSE) \cite{Ref24}. Here, the proposed veMSE was utilized to exhibit an improved performance of characterizing different dynamical complexity of wind regimes.

\begin{figure}[htp]
    \centering
    \includegraphics[width=5in]{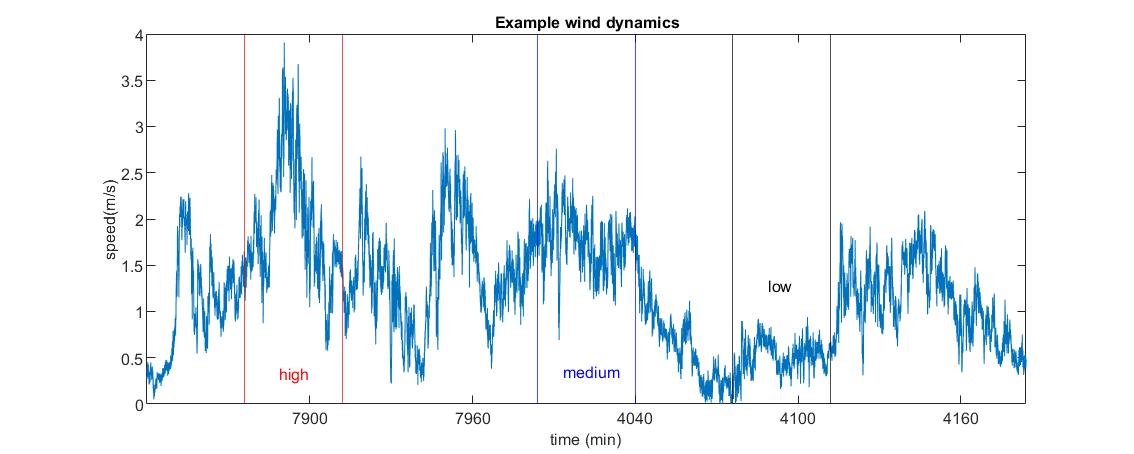}
    \caption{Magnitude of the wind signal. The wind segments are defined as low, medium, and high regimes.}
    \label{fig:wind}
\end{figure}

The applied database of wind was recorded by 3D ultrasonic anemometers at a sampling frequency of 50 Hz. The recording process was implemented in the Institute of Industrial Science of the University of Tokyo. Three channels were obtained and they are in east-west, north-south, and vertical direction respectively. The wind regimes containing different dynamics were defined as low, medium, and high by the magnitude of wind speed as examples shown in Figure. \ref{fig:wind}. The parameters involved in veMSE analysis were set to $m$ = 2, $L$ = 1, $r = 0.2\times tr(S)$, and $N$ = 3000. Illustrations of entropy results were averaged over  10 trials for each channel and exhibited as error-bar. In addition, shuffled wind data sets were also tested and compared with obtained wind dynamics, which were generated from the recorded wind signals but with a random order. In this case, the possible correlations within signals were broken down but the statistical properties were preserved.

\begin{figure}[htbp]
        \centering 
         \subfigure[]{
        \includegraphics[width=2.2in]{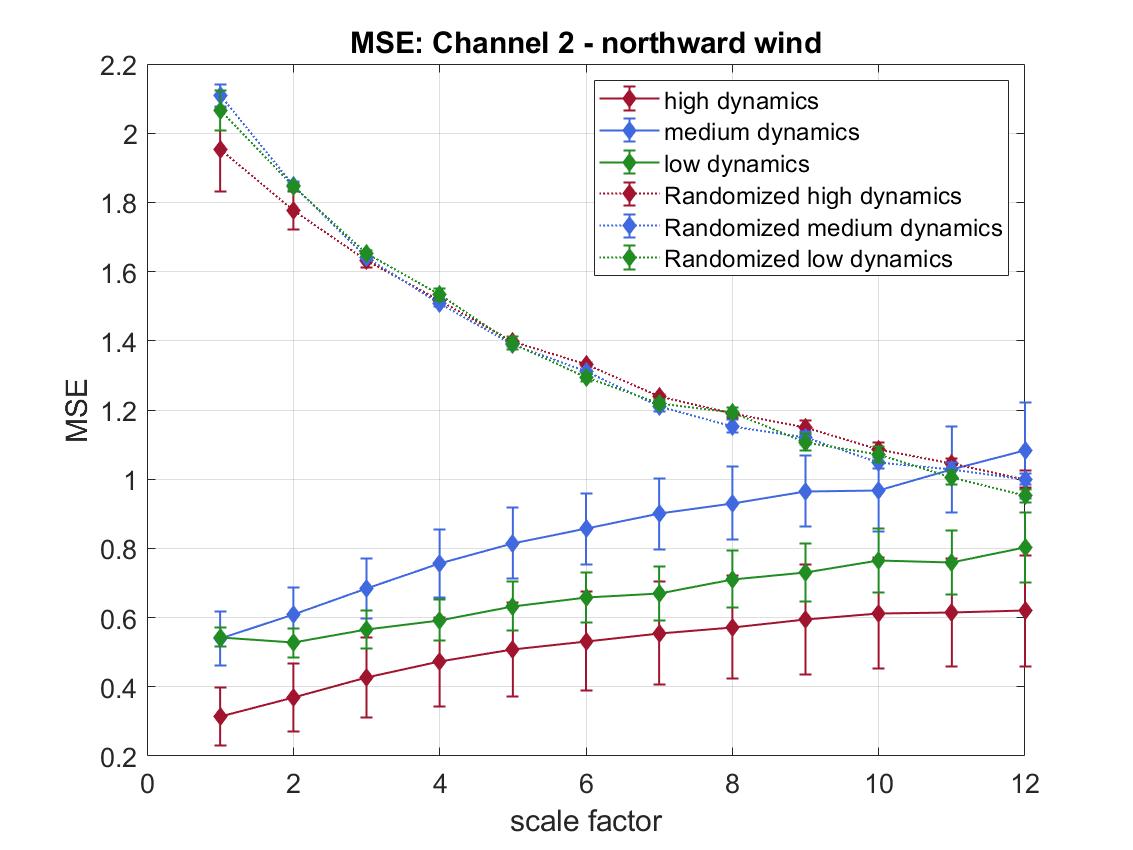}
        \label{fig:MSE_wind}
        }
        \subfigure[]{
        \includegraphics[width=2.2in]{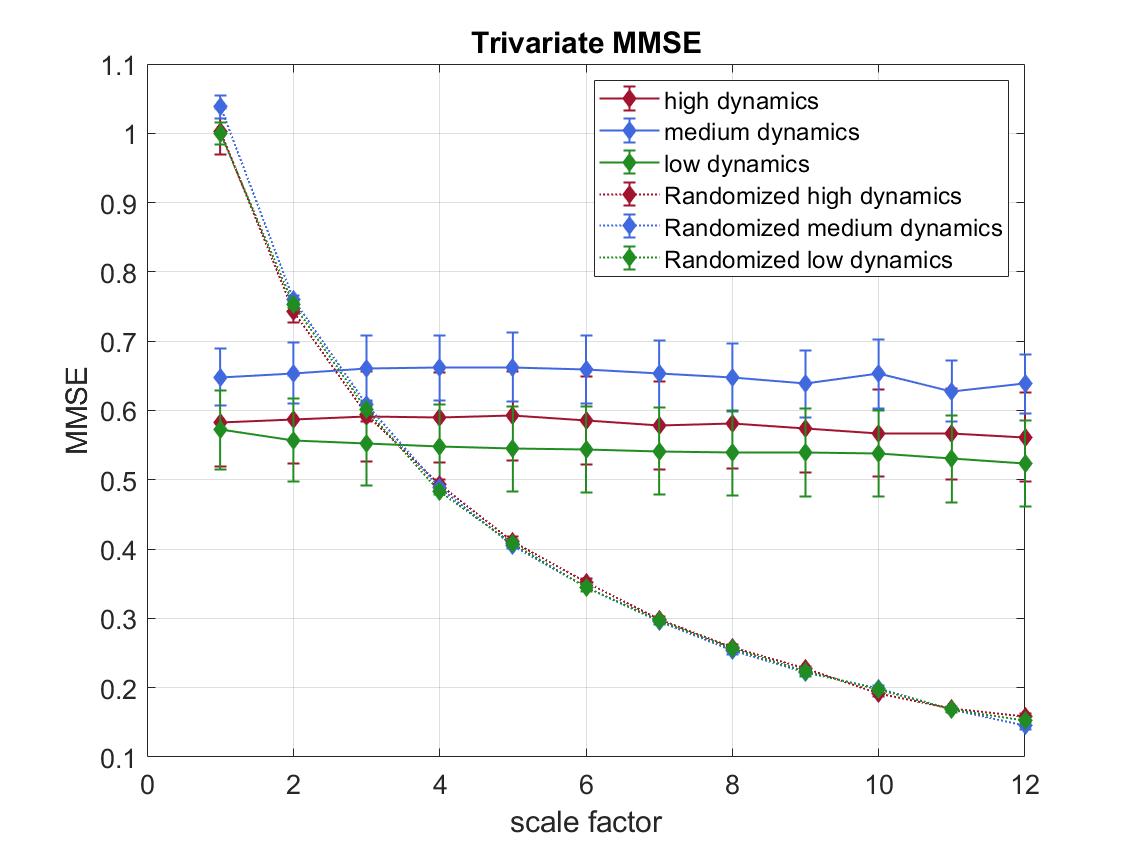}
           \label{fig:MMSE_wind}}
            \subfigure[]{
        \includegraphics[width=2.2in]{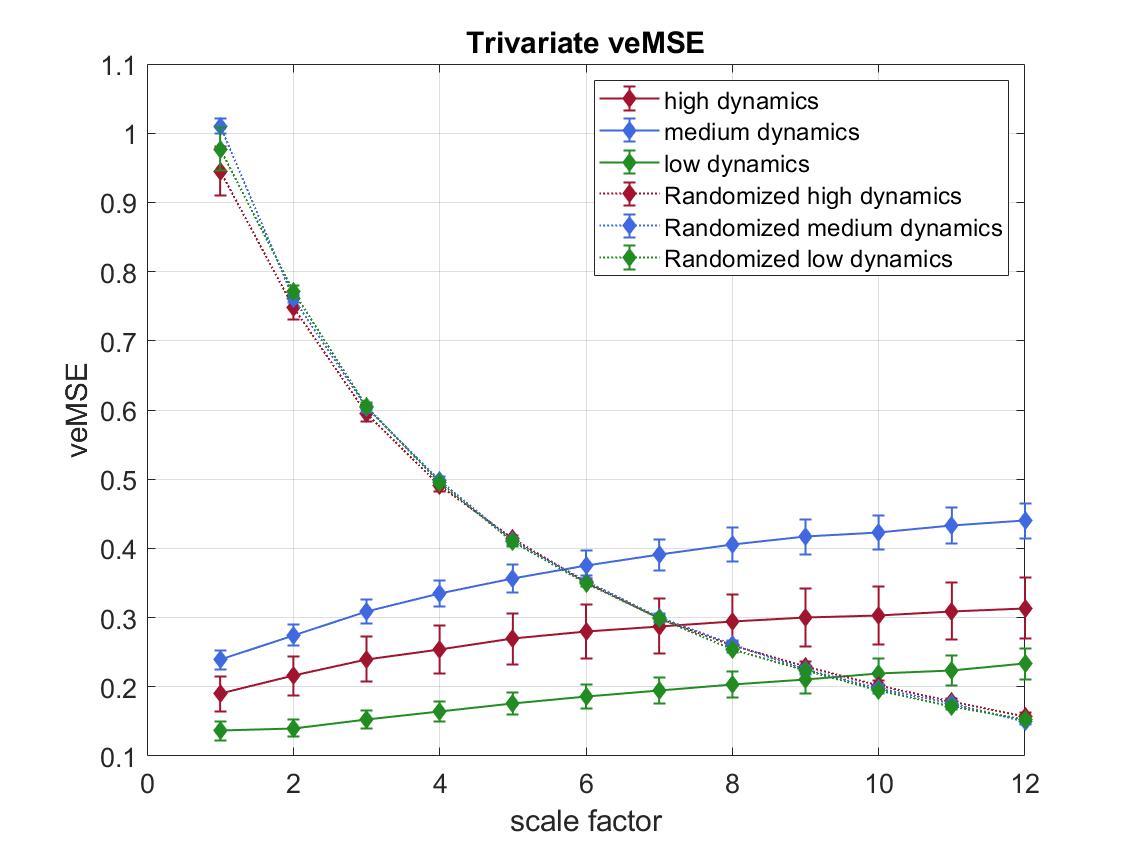}
           \label{fig:veMSE_wind}}
        \caption{Results of a) standard single-variate MSE, b) standard multi-channel MMSE, c) proposed veMSE based on wind dynamics and randomized dynamics.}
         \label{fig:wind_entropy}
\end{figure}

Before the comparison between veMSE and MMSE, univariate Multiscale Sample Entropy (MSE) was firstly applied to give an insight into the complexity of the dynamical system to be processed. As shown in Figure.\ref{fig:MSE_wind}, the randomized data sets exhibit a white noise-like behaviour in terms of structural complexity, which is expected as the correlation was destroyed while shuffling. The MSE results of wind dynamics assigned the highest complexity to the medium regime which is expected since the medium wind speed hold the fewest constrains. Yet, the complexity of low wind regime was quantified higher than the high regime which violates the intuition, as the high wind regime also contains components with medium speed. Besides, each wind dynamics exhibited lower complexity than the their randomized series, wrongly resulting to the absence of long-term correlations within wind data sets.

The same trails were analyzed by Multivariate Multiscale Sample Entropy as shown in Figure.\ref{fig:MMSE_wind} which revealed the relation of the three wind regimes in a expected way that mild wind exhibited the highest complexity, followed by the high and low dynamics regimes. Further, all the three wind regimes were able to present long-range correlation higher than their surrogate data sets. However, overlapped area can be found across all the scales. While in Figure.\ref{fig:veMSE_wind}, observe, as desired, the long-range correlation of different wind speed was able to be detected resulting from the proposed veMSE with more apparent separation on basis of the correct analysis compared to MSE and MMSE.

\subsection{Physiological Database}
Now, physiological data sets were involved to exhibit the performance of complexity quantification based on entropy analysis. The physiological signals utilized is Fantasia database \cite{Ref40} that includes R-R intervals (RRI) and interbreath interval (IBI) extracted from electrocardiograph (ECG) and respiration signals respectively. The structure of long-term correlation in heart rate variability and respiratory dynamics have been examined by traditional method as Detrended Fluctuation Analysis (DFA) \cite{Ref277} and standard MMSE \cite{Ref24}. 

Fantasia database were recorded from 20 young people (at the age of 21-34) and 20 elderly participants (ageing from 68-85) who were rigorously screened for 120 minutes. 10 subsets out of them were chosen in each group. The ECG and respiration signals were recorded with sample frequency at 250 Hz. Then, RRI and IBI signals were extracted and aligned. The surrogate series with randomised order were analyzed along with RRI and IBI signals in each methodology to reveal the effectiveness of the complexity. Parameters in entropy analysis were set to $m$ = 2, $L$ = 1, $r = 0.15\times tr(S)$, and $N$ = 4000.

\begin{figure}[htbp]
\centering
\begin{minipage}[t]{3.3in}
\centering
\includegraphics[width=3.1in]{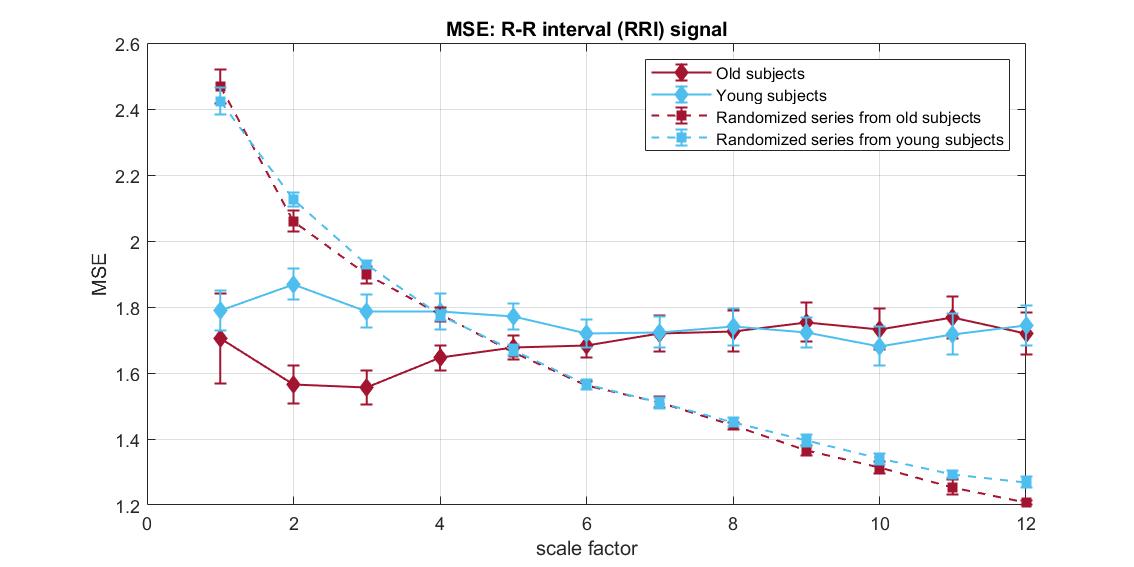}
\caption{Operation of Univariate Multiscale Sample Entropy based on R-R interval.}
\label{fig:MSE_RRI}
\end{minipage}
\hspace{0.1in}
\begin{minipage}[t]{3.3in}
\centering
\includegraphics[width=3.1in]{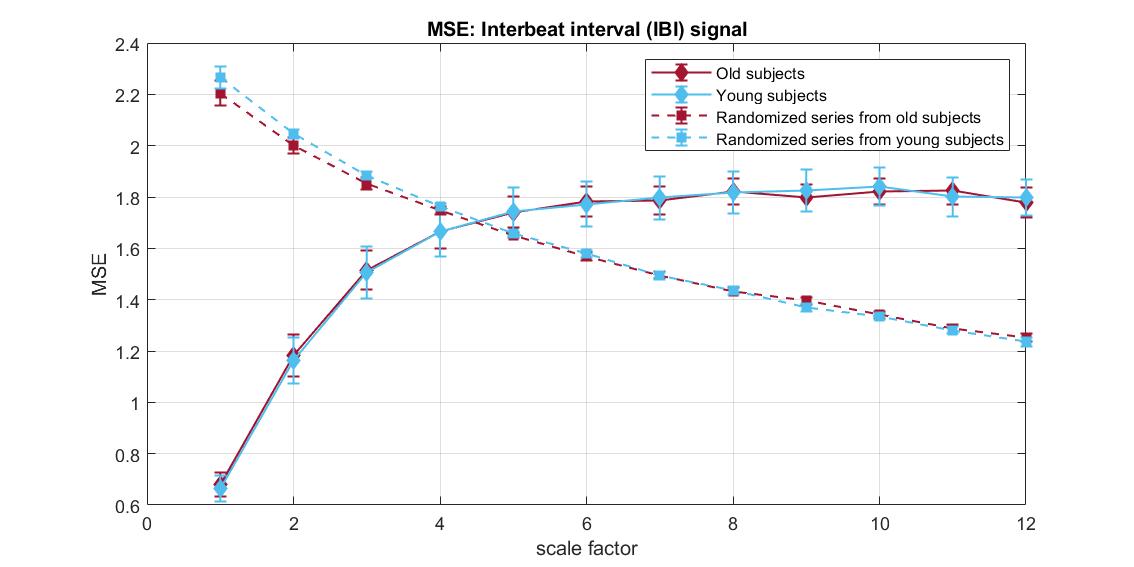}
\caption{Operation of Univariate Multiscale Sample Entropy based on interbreath interval (IBI).}
\label{fig:MSE_IBI}
\end{minipage}
\end{figure}

In the first place, to give an overall view of the signals in each channel. Univariate Multiscale Sample Entropy was applied to single channel separately as exhibited in Figure.\ref{fig:MSE_RRI} and Figure.\ref{fig:MSE_IBI}. As Complexity Loss Theory \cite{Ref254} stated, the adaptive capacity of bio-systems is damaged by disease and aging. The complexity of RRI and IBI signals recorded from young people is supposed to be higher than that from elderly subjects when considering long-term correlation. In respect of RRI in short-term correlation, MSE exhibited a correct estimation under certain low scales. However, both MSE on basis of RRI and IBI are unable to reveal the correct relation between the dynamics of elder individuals and young people in a long-range term.

\begin{figure}[htbp]
\centering
\begin{minipage}[t]{3.3in}
\centering
\includegraphics[width=3.1in]{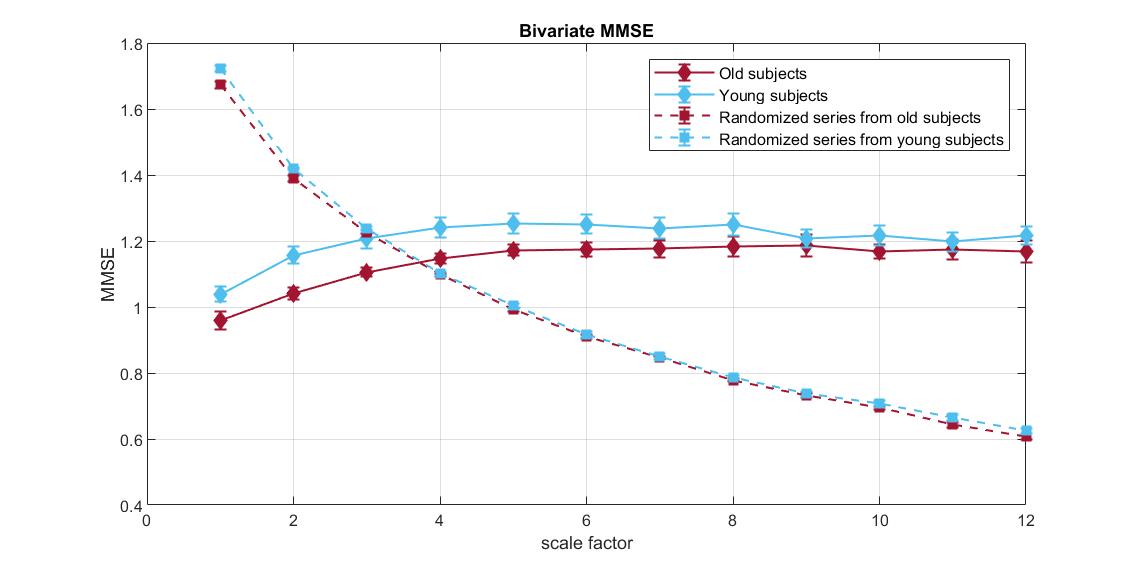}
\caption{Operation of Multivariate Multiscale Sample Entropy.}
\label{fig:MMSE_Fan}
\end{minipage}
\hspace{0.1in}
\begin{minipage}[t]{3.3in}
\centering
\includegraphics[width=3.1in]{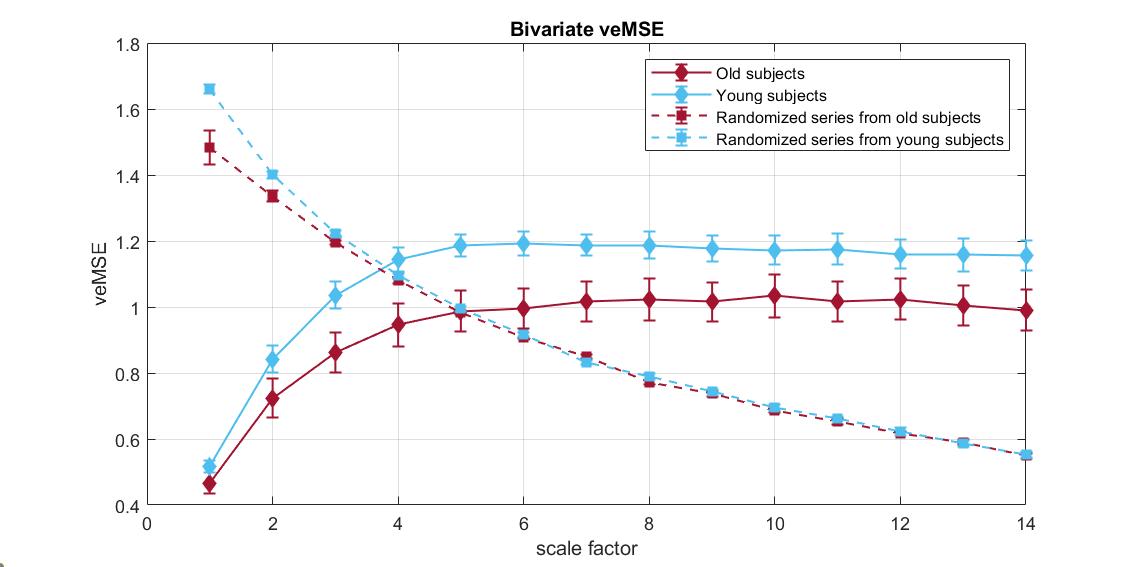}
\caption{Operation of Variational Embedding Multiscale Sample Entropy.}
\label{fig:veMSE_Fan}
\end{minipage}
\end{figure}

Next, observed in Figure.\ref{fig:MMSE_Fan}, MMSE was applied to the bivariate channels. The resulted error bar indicated the higher dynamical complexity in young participants than that in elderly individuals which conformed the complexity loss theory with aging \cite{Ref254}. Moreover, both physiological data sets showed higher long-range correlation than the randomized surrogate series in an expected way, yet as scale increases, overlapping area can be observed due to the lack of sample length after coarse graining process. While in contrast to MMSE, performance of veMSE in Figure.\ref{fig:veMSE_Fan} remained the correct and clear estimation but with higher stability particularly in large scales. In practical scenarios, the size of recorded signals is mostly limited. Hence, veMSE can better facilitate the analysis of physical and physiological signals when identifying dynamical difference based on nonlinear features.

\section{Conclusion}

To conclude, the Variational Embedding Multiscale Sample Entropy (veMSE) method has been introduced for enhanced complexity analysis of real-world data. The results of this paper indicate that veMSE is capable of exhibiting the complex features of the system in large scale with higher embedding dimension compared to MMSE and veMDE. Besides, the utilization of multivariate analysis via veMSE could guarantee an improvement in contrast to single-variate analysis regardless of the quality of the recorded signals in sub-channels. Moreover, veMSE shows a strong noise robustness feature and meanwhile, the time consumption of veMSE is lower than MMSE under the same conditions as well, and this improvement is particularly apparent as the increase of available channels. The higher calculation efficiency is of high potential interest to apply entropy analysis into scenarios where need real-time processing or synchronized monitor in the future application. Nevertheless, it should be noted that this method is restricted to amplitude-based distance. Future research can be intended to put focus on angular distance-based associated with variational embedding dimension methodology.

\appendices

\section{Algorithm of Multivariate Multiscale Sample Entropy}

\begin{algorithm}[htb]

    \caption{Multivariate Multiscale Sample Entropy }
    \label{MMSE}
    \begin{algorithmic}
\Statex The steps of standard Multivariate Multiscale Sample Entropy are given below. For a multi-variate data set $\{x_{c,i}\}_{i=1}^N,\,1\leq c\leq P$ with length of $N$ and number of channel $P$. Manually selected parameters are the embedding dimension ($M = [m_1, m_2,\dots, m_P]$), tolerance ($r$), time delay ($L = [l_1, l_2,\dots, l_P]$) and scale factor ($\tau$):
    \begin{enumerate}
        \item Normalize the original multi-variate data sets by subtracting the mean and dividing by the standard deviation.
        
        \item Perform Coarse Graining Process to obtain the scaled multi-channel time series $\{y_{c,i}^{(\tau)}\}_{j=1}^{N/\tau}$ following the equation\newline
                \begin{equation}
                y_{c,i}^{(\tau)}(j) = \frac{1}{\tau}\sum^{j+\tau/2-1}_{i = j-\tau/2 - 1}{x_{c}(i)},\quad 1\leq j\leq \frac{N}{\tau},\,c = 1,2,\dots,P
                \nonumber
                \end{equation}
        \item Form the Composite Delay Vectors (CDV) $\textbf{Y}_M(i)$ according to $M$ and $L$ in a form as
                \begin{align*}
                \textbf{Y}_M(i) =[& y_{1,i},y_{1,i+l_1},\dots,y_{1,i+(m_1-1)l_1)},\\
                                 & y_{2,i},y_{2,i+l_2},\dots,y_{2,i+(m_2-1)l_2)},\\
                                 &\qquad \qquad \qquad \vdots \\
                                 & y_{c,i},y_{c,i+l_c},\dots,y_{c,i+(m_c-1)l_c)},]
                 \end{align*}
        \item Compute the similarity for all pairwise CDVs ($Y_M(i)\,\&\, Y_M(j)$) based on Chebyshev distance as
                \begin{equation}
                D(i,j) = max\{ Y_M(i+k)-Y_M(j+k)||\, 0\leq k\leq ((\sum_{c=1}^{P}{m_c})-1),i\neq j\} 
                \nonumber
                \end{equation}
        \item Calculate the number of matching patterns defined as similar pair $B(i)$ that satisfies the criterion $D(i,j)\leq r$.
        \item Compute the local probability $C(i)$ and global probability $\Phi(i)$ of $B(i)$ by \newline
                \begin{equation}
                C(i) = \frac{B(i)}{N-n-1},\quad \Phi=\frac{\sum_{i=1}^{N-n}{{C(i)}}}{N-n},\,n = max(M)*max(L)
                \nonumber
                \end{equation}
        \item Repeat Steps 3-6 with modified embedding dimension as ($m_c$+1) and obtain the updated global probability as\newline
                \begin{equation}
                \Phi^*=\frac{\sum_{i=1}^{N-n}{{C^*(i)}}}{N-n},\, n = max(M^*)*max(L)
                \nonumber
                \end{equation}
          Recall that there are $P$ ways to increase the embedding dimension and the modified global probability, $\Phi^*$, is the averaged result.
           \item Multivariate Multiscale Sample Entropy is defined as\newline
                    \begin{equation}
               MMSE= -\ln{[\frac{\Phi^*}{\Phi}]}
                \nonumber
                \end{equation}
    \end{enumerate}

    \end{algorithmic}
\end{algorithm}

\section*{Acknowledgment}
We wish to thank Institute of Industrial Science of the University of Tokyo for providing the wind data sets used in this article.

\ifCLASSOPTIONcaptionsoff
  \newpage
\fi



%
\bibliographystyle{ieeetr}
\bibliography{bibtex/ref}

%

\end{document}